\documentclass{SciPost}
\hypersetup{
    colorlinks,
    linkcolor={red!50!black},
    citecolor={blue!50!black},
    urlcolor={blue!80!black}
}

\usepackage[bitstream-charter]{mathdesign}
\usepackage{graphicx}
\DeclareSymbolFont{usualmathcal}{OMS}{cmsy}{m}{n}
\DeclareSymbolFontAlphabet{\mathcal}{usualmathcal}
\fancypagestyle{SPstyle}{
\fancyhf{}
\lhead{\colorbox{scipostblue}{\bf \color{white} ~SciPost Physics}}
\rhead{{\bf \color{scipostdeepblue} ~Submission }}

\fancyfoot[C]{\textbf{\thepage}}
}

\begin{document}
\pagestyle{SPstyle}

\begin{center}{\Large \textbf{\color{scipostdeepblue}{
Fate of emergent Hilbert space fragmentation and information insulation upon softening of hard dual constraints at PXP-chain junctions\\
}}}\end{center}

\begin{center}\textbf{
Akshay Panda\textsuperscript{1$\star$},
Anwesha Chattopadhyay\textsuperscript{1$\dagger$}
}\end{center}

\begin{center}
{\bf 1} Department of Physics, School of Mathematical Sciences\\
Ramakrishna Mission Vivekananda Educational and Research Institute\\
PO Belur Math, Dist. Howrah 711202, West Bengal, India
\\

$\star$ \href{mailto:email1}{\small akshay.panda@gm.rkmvu.ac.in}\,,\quad
$\dagger$ \href{mailto:email2}{\small anwesha.phy@gm.rkmvu.ac.in},\\
$\dagger$ \href{mailto:email2}{\small anweshachattopadhyay@yahoo.com}
\end{center}

\section*{\color{scipostdeepblue}{Abstract}}
\textbf{
We demonstrate emergent Hilbert-space fragmentation in a composite chain of PXP and dual-PXP subchains separated by junctions with simultaneous blockade and anti-blockade projection constraints. The Hilbert space shatters into disjoint Krylov fragments growing exponentially with junction number. Freezing does not require a unique allowed junction state: multiple configurations remain locally allowed, but the dual constraints remove dynamical links between them. Each fragment is labeled by frozen junction configurations and contains compatible basis states. This fragmentation produces dynamical caging of quantum information between neighboring junctions and novel states with partially thermal and athermal regions. Softening the junction by admitting other local states with variable nonzero diagonal amplitudes destroys the direct-sum block structure and fragmentation. Nevertheless, information remains dynamically insulated and inhomogeneous thermal properties persist up to an onset time scaling with the diagonal energy penalty of the previously restricted junction states, followed by longer prethermal behavior. Energy-level statistics in each symmetry-resolved sector evolve from Poissonian in the hard-junction limit toward Wigner-Dyson behavior upon softening, with finite-size non-monotonicity. This signals the breakdown of the tensor-sum structure yielding pairwise sums of chaotic subchain spectra, restoring chaotic thermalization. In the hard-junction case, we find a chirally protected zero-energy mode with density peaks at physical edges and near the junction, pinned to zero under chirality-preserving disorder. The tensor-product eigenfunction structure also enhances the number of nonzero-energy scar states. Finally, we discuss routes to these junction architectures in programmable Rydberg atom platforms using optical tweezers, local addressing, and facilitation techniques.}



\vspace{10pt}
\noindent\rule{\textwidth}{1pt}
\tableofcontents
\noindent\rule{\textwidth}{1pt}
\vspace{10pt}


\section{Introduction}
Recently, there has been an upsurge in devising ways to store quantum information and transport it between distant nodes without decoherence~\cite{QI1,QI2,QI3,QI4,QI5,QI6}. The major problem in storing and transporting quantum information is that a qubit of information in the initial state very easily entangles with the thermal bulk states and undergoes catastrophic decoherence. To this end, topological architectures involving Majorana edge modes, which are topologically protected by a macroscopic bulk gap, are being designed to store quantum information robustly~\cite{top1,top2,top3,top4,top5,top6,top7}. Alternatively, quantum information can be shielded dynamically by breaking ergodicity in isolated quantum many-body systems. There are several paradigms where ergodicity can be violated; for example, in many-body localized systems~\cite{MBL1,MBL2,MBL3}, local integrals of motion arise because of the interplay of strong disorder and interactions. Integrable systems~\cite{INT1,INT2} have an extensive number of conserved quantities which trap the system in some conserved sector and do not allow it to thermalize. Recently, the focus has shifted to disorder-free mechanisms of ergodicity breaking such as Hilbert space fragmentation~\cite{HSF1,HSF2,HSF3,HSF4,HSF5}, where disconnected sectors are formed in the Hilbert space. These are not governed by any underlying global symmetry, but arise due to kinematic constraints in the system. Systems can also show weak ergodicity breaking due to the existence of an algebraic number of scar states~\cite{PXP2,PXP3,Rydexpt}, which are athermal and show sub-volume entanglement entropy in contrast to the bulk of the spectrum in which they are embedded. The physics of these athermal scars was experimentally observed in a 51-atom Rydberg array~\cite{Rydexpt} where atoms in the excited or Rydberg state were suppressed on the nearest-neighbor sites due to strong repulsive interaction. Using optical tweezers, the neutral atoms were arranged in a 1D array and using addressing beams the initial $Z_{2}$ state was prepared consisting of atoms in alternating Rydberg ground and excited states. This isolated system was allowed to evolve unitarily and it was observed that the system avoided thermalization and exhibited long-lived coherent oscillations between $Z_{2}\leftrightarrow \bar{Z}_{2}$ where the $\bar{Z}_{2}$ state is the state with  $0\leftrightarrow 1$ in the $Z_{2}$ state. This experiment was theoretically explained by the $PXP$ model~\cite{PXP1,PXP2,PXP3} in 1D where spin-$\frac{1}{2}$ particles are flipped subject to the constraint that no two up-spins occupy nearest-neighbor sites. It was found that the states like $Z_{2}$ had a large overlap with these scar states. Therefore, these special initial states retain memory and weakly violate eigenstate thermalization hypothesis (ETH)~\cite{ETH1,ETH2,ETH3}.

If we consider an isolated quantum system, the full system remains in a pure state and evolves unitarily, so the information about the initial state is never truly lost. Then the question arises as to why we talk about thermalisation in an isolated quantum system. If we consider only a small subsystem \(A\), although the density matrix \(\rho_{\mathrm{total}}\) of the full system remains pure, tracing over the rest of the system \(\bar A\) gives a reduced density matrix \(\rho_A=\mathrm{Tr}_{\bar A}\textit{ } \rho_{\mathrm{total}}\), which can become mixed due to entanglement with the rest of the system. In a generic thermalising system, at long times this reduced density matrix approaches the corresponding thermal reduced density matrix, with an effective temperature fixed by the conserved energy density of the full system. Thus, the rest of the system acts as an effective bath for the subsystem, even though the total system is isolated. In this sense, local information can be lost to the many-body bulk while the total information of the full system is still preserved. The question we ask in this paper is complementary: how can a subsystem avoid thermalising with the rest of the bulk and thereby retain its local information? Our main objective is to look for ways in which a subsystem can be dynamically isolated from the rest of the system. This, as we will show, can be achieved by caging information between suitably engineered junctions. In this paper, firstly, we devise a novel mechanism to cage quantum information by constructing a model which shows emergent Hilbert space fragmentation. The model consists of a chain that contains consecutive $PXP$ and $d-PXP$ subchains with junctions where both blockade and anti-blockade constraints are imposed. It should be noted that the physics we seek from this model remains valid even if we have ordinary $PXP$ subchains glued by junctions with such dual constraints. However, we focus on consecutive \(PXP\) and \(d\)-\(PXP\) subchains because this architecture provides a natural theoretical setting in which blockade and anti-blockade constraints meet at an interface. Due to the dual nature of constraints at the junction, the dynamical linkage between the allowed states at the junction is broken leading to an emergent freezing of the junction states. This leads to the fragmenting of the Hilbert space where each fragment is labeled by the state of junctions in the configuration space. This immediately results in the dynamical signature of information localization between neighboring junctions. Additionally, it is also possible to engineer novel states which are partially thermal and partially athermal. It is to be noted that this confinement has two aspects: Hilbert-space fragmentation restricts the dynamics to a finite Krylov sector, thereby limiting the set of Fock states over which the initial information can spread, while the tensor-product decomposition induced by frozen junctions additionally confines this spreading to a finite spatial region of the chain. Since the Hilbert-space fragmentation is parametrized by the frozen junction configurations, the resulting Fock-space and real-space localization are intrinsically intertwined. 

Secondly, we explore the physics of this model when the hard constraints on the junction are relaxed to different degrees leading to a soft constraint version. It is interesting to focus on this variant primarily due to the expected experimental ease of preparing such a system with a soft junction rather than a hard junction. Softening the junction  kills the exact Hilbert space fragmentation because the hard blockade and anti-blockade constraints are relaxed at the junction bond, allowing all four states on the junction bond which are dynamically coupled through nonzero flip amplitudes. However, the amplitude of information tunneling across the bond depends on the parameter \(\Delta\), which sets the finite energy penalty for the previously forbidden junction states. \(\Delta\) tunes the effective energy barrier at the junction and hence controls the tunneling amplitude, or equivalently how rapidly the previously allowed junction states become connected through the previously forbidden intermediate states. When $\Delta \rightarrow \infty$, we are in the hard-junction limit. This parameter also controls the information-onset timescale over which information remains effectively insulated between junctions and distinct thermal properties of the regions can be preserved. This timescale is followed by a long prethermal window over which the subchains get progressively entangled with each other. Therefore, a soft junction with a reasonably high $\Delta$ can be an experimental playground for dynamically isolating information and retaining it up to a timescale which is long but upper bounded by the prethermal timescale.

We also establish that in the symmetry-resolved sector of the Hamiltonian, the level repulsion statistics go from the Poissonian hard-junction limit to Wigner-Dyson as the junction is gradually softened by reducing $\Delta$, apart from finite-size effect of intermediate non-monotonicity. However, the $\Delta=0$ point again shows Poissonian level repulsion statistics. The tensor-sum structure of the disjoint Hamiltonians in the hard-junction limit leads to sums of the independent chaotic spectra of the sub-\(PXP\) or $d-PXP$ chains. For finite $\Delta$, the tensor-sum structure is destroyed leading to the chaotic spectra of thermalizing chains akin to homogeneous $PXP$ chain. At $\Delta=0$, when the kinetic couplings of the junction states are equal, the tensor-sum structure of the Hamiltonian is restored. Further, in the case of hard junction, the tensor product structure of the eigenfunctions also leads to the increase in the number of nonzero energy scar eigenstates as compared to a continuous $PXP$ chain of the same length. In addition, we find zero-energy modes protected by chiral symmetry, whose spatial profiles exhibit pronounced peaks near the physical boundaries and the hard junctions. Within the zero-mode subspace, we can also construct states with anomalously low entanglement entropy, providing an athermal zero-mode analogue of scarred states. With the advent of programmable Rydberg-atom platforms using optical tweezers, addressing beams, and facilitation techniques, these features can be explored experimentally and may provide a promising route toward applications in local quantum-information isolation, storage, and retrieval.

\section{Model}
We join an open boundary $PXP$ chain with an equal-length dual $PXP$ chain, each of which is either even or odd in length. On the left half of the chain, two neighboring up-spins are not allowed whereas on the right half of the chain, two neighboring down-spins are not allowed. In the hard-core boson language, this  means that the probability of two bosons sitting next to each other is zero on the left half, whereas the probability of two holes sitting next to each other is zero on the right half of the chain. 
 The Hamiltonian of the $PXP$- dual $PXP$ chain is defined as, 

\begin{align}
\mathcal{H}
={}&
\sum_{i=2}^{L/2-1}
P^{i-1}_{l}\sigma_{X}^{i}P^{i+1}_{l}
+
\sum_{i=L/2+2}^{L-1}
P^{i-1}_{r}\sigma_{X}^{i}P^{i+1}_{r}
+
H^{J}_{L/2,L/2+1}
+
\sigma^{1}_{x}P^{2}_{l}
+
P_{r}^{L-1}\sigma^{L}_{x}.
\label{eq:composite_hamiltonian}
\end{align}
  
where $P^{i}_{r,l}=\dfrac{1\pm \sigma^{i}_{z}}{2}$ and 
$H^{J}_{L/2,L/2+1}=\Delta_{00}|00\rangle\langle00|+\Delta_{11}|11\rangle\langle 11|+\epsilon_{00}(P_{l}^{L/2-1}|10\rangle\langle00|+P_{r}^{L/2+2}\\|00\rangle\langle01|+ h.c.)+\epsilon_{11}(P_{r}^{L/2+2}|10\rangle\langle11|+P_{l}^{L/2-1}|11\rangle\langle01|+ h.c.)$. The following sets of parameter values yield different physical properties,

\begin{itemize}
    \item $\Delta_{00,11}=0, \epsilon_{00,11}\neq 0$ are the parameter values at which all $2^{2}$ states at the junction i.e., $\{00,11,01,10\}$ are allowed at the junction Hilbert space. The flipping terms in the case of $\epsilon_{00}=\epsilon_{11}$ at the junction sites are decoupled $\sigma_{x}$ operators which are mutually non-interacting and hence this is an example of a ``free" bridge bond in a constrained sea. In this case, the Hamiltonian is a tensor-sum of the left and right $PXP$ and $d-PXP$ chains which are dynamically disjoint. This is a trivial example of a kinetically decoupled junction. For $\epsilon_{00} \ne \epsilon_{11}$, kinetically induced interaction terms such as $\sigma_{x}^{L/2}\sigma_{z}^{L/2+1}$ and $\sigma_{z}^{L/2}\sigma_{x}^{L/2+1}$ arise that entangle the junction spins.

 \item $\Delta_{00}\rightarrow \infty, \Delta_{11}=0, \epsilon_{00,11}\neq 0$ or $\Delta_{11}\rightarrow \infty, \Delta_{00}=0, \epsilon_{00,11}\neq 0$ corresponds to the case where one of the constraints of not allowing $00/11$ at the junction is enforced, thereby resulting in a  $2^{2}-1$ dimensional local Hilbert space at the junction. This is an example of a ``soft" junction where the dynamical two-step process $01 \rightarrow 00 \rightarrow 10$ or $01 \rightarrow 11 \rightarrow 10$ is active. In this limit, the constrained space $\{11/00,01,10\}$ creates an effective constraint-driven interaction (conditional kinetic terms) between the two junction sites which entangles them. The Hilbert space shrinks as compared to the above case.

    \item $\Delta_{00,11}\rightarrow \infty$, $\epsilon_{00,11} \neq 0$ corresponds to the exact analytical ``hard" junction point where the $00$ and $11$ nearest-neighbor configurations on the junction bond are infinitely costly and therefore are not allowed in the junction Hilbert space. This makes the flipping terms corresponding to $\epsilon _{00,11}$ ineffective. As a result of the dual constraints at the junction, the two allowed states $|01\rangle$  and $|10\rangle$ become dynamically disconnected because the two-step processes of this single-flip Hamiltonian like $01 \rightarrow 00\rightarrow 10$ or $01 \rightarrow 11 \rightarrow 10$ are no longer possible. The Hilbert space projector at the junction then simply becomes $\mathcal{P}^{J}_{L/2,L/2+1}=1 - \frac{1}{4}(1+\sigma^{L/2}_{z})(1+\sigma^{L/2+1}_{z})- \frac{1}{4}(1-\sigma^{L/2}_{z})(1-\sigma^{L/2+1}_{z})$ which in the hard-core boson language corresponds to $\mathcal{P}^{J}_{L/2,L/2+1}=I-|11\rangle\langle11|-|00\rangle\langle 00|=|01\rangle\langle01|+|10\rangle\langle10|$ from the completeness of the Hilbert space at the two junction sites. The effective Hilbert space on which the Hamiltonian in Eq.~\eqref{eq:composite_hamiltonian} acts ( in the hard-junction limit) can therefore be obtained by acting $\mathcal{P}_{l}\mathcal{P}_{J}\mathcal{P}_{r}$ on the unrestricted $2^{L}$ Fock space, where $\mathcal{P}_{l}=\prod_{i=1}^{L/2-1} (I-|1_{i}1_{i+1}\rangle\langle 1_{i}1_{i+1}|)$ and $\mathcal{P}_{r}=\prod_{i=L/2+1}^{L-1} (I-|0_{i}0_{i+1}\rangle\langle 0_{i}0_{i+1}|)$ are the Hilbert space projectors on the bonds on the left and right half of the chain. This model is fragmented into two sectors labeled by the state of the junction sites i.e., $01$ and $10$ due to an emergent \(\mathbb Z_2\) symmetry of the junction. We can define a local operator $T=|01\rangle\langle 01|-|10\rangle\langle 10|$ defined on the junction sites which has eigenvalues $\pm 1$ in the $01$ and $10$ sectors respectively. Since \([H,T]=0\), \(T\) acts as an emergent local \(\mathbb Z_2\) charge distinguishing the two fragmented sectors. The $01$ and $10$ sectors are dynamically disconnected and there are no off-diagonal terms which connect $01 \leftrightarrow 10$ at the junction sites. It is to be noted that as we increase the multiplicity of the defect junctions, the kinematically disconnected Krylov spaces increase exponentially in the number of defect junctions in the chain.  Consequently, for a finite density of junctions, \(n\propto L\), the number of fragments also grows exponentially with system size, as in strongly fragmented systems. For a fixed number of junctions, however, the number of fragments remains finite in the thermodynamic limit. If we have $n$ junctions, there are exactly $2^n$ disconnected Krylov subspaces. In figure~\ref{fig1}(a), we show a $PXP-d$-$PXP-PXP-d$-$PXP$ chain of length $L=16$ with $n=3$ junctions, for which in figure~\ref{fig1}(b) we show the connectivity graph among the Fock states in the Hilbert space.

    \begin{figure}[h!]
    \centering
    \includegraphics[width=0.48\linewidth,height=5.8cm]{ 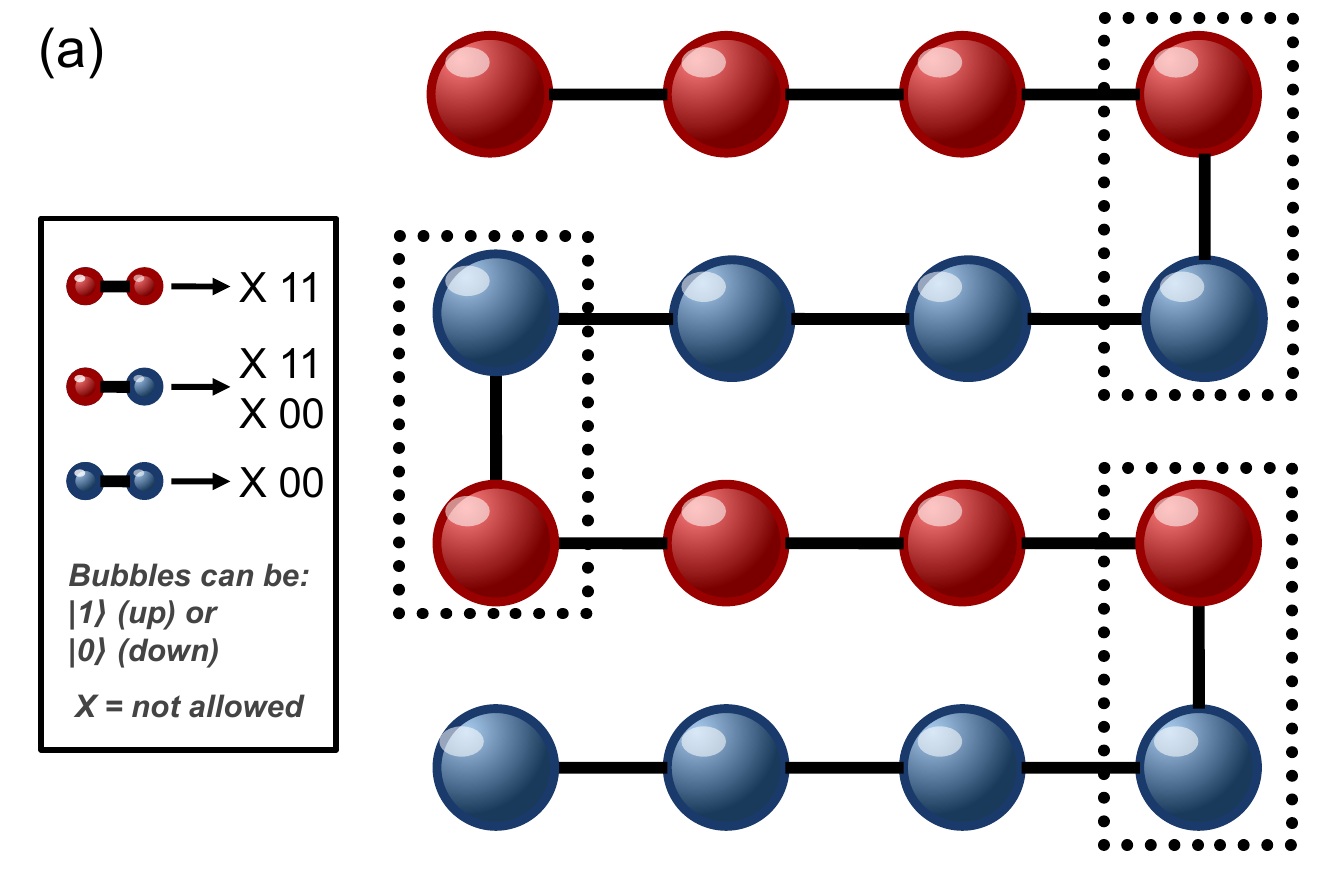}
    \includegraphics[width=0.48\linewidth,height=6.0cm]{ 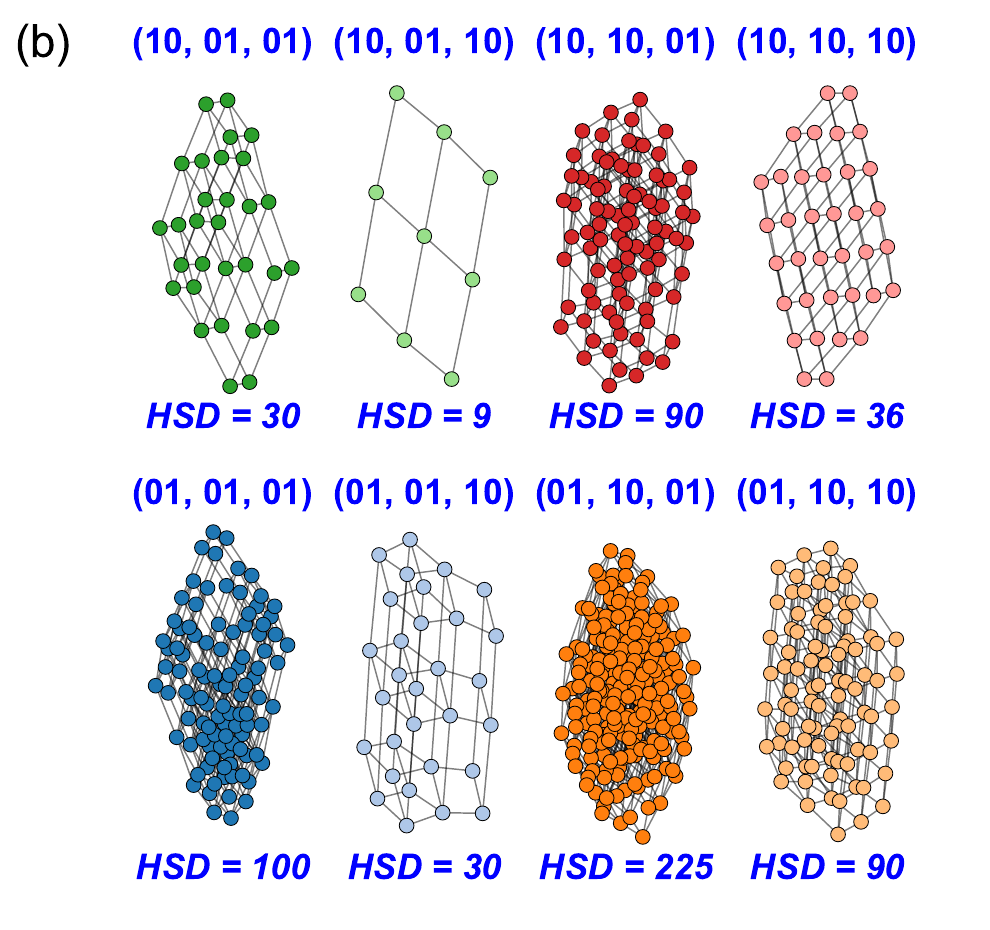}
    \caption{Panel $(a)$ shows an open $PXP-d$-$PXP-PXP-d$-$PXP$ chain of length, $L=16$ with three hard junctions each of which does not allow $00$ and $11$. Panel $(b)$ shows all the Hilbert space fragments of varying Hilbert space dimensions ($HSD$) for the chain in panel (a).}
    {\label{fig1}}
\end{figure}

    \item For $\Delta_{00,11}< \infty, \epsilon_{00,11}\neq 0$, the Hamiltonian is not necessarily traceless like the previous cases. This eliminates the $E\leftrightarrow-E$ symmetry in the eigen spectrum. Further, even though the junction Hilbert space is now again unconstrained like the first case, the finite diagonal elements introduce interaction $\dfrac{\Delta_{00}}{4}(1-\sigma^{L/2}_{z})(1-\sigma^{L/2+1}_{z})+\dfrac{\Delta_{11}}{4}(1+\sigma^{L/2}_{z})(1+\sigma^{L/2+1}_{z})$ between the junction spins and makes it ``soft".  The Hilbert space graph has the same connection as that of the first case and consequently, there is no Hilbert space fragmentation in this case. It is to be noted that all the numerical data that will be presented in the following sections will correspond to $\Delta_{00}=\Delta_{11}=\Delta$ and $\epsilon_{00}=\epsilon_{11}=1$.

\end{itemize}

\begin{equation}
\resizebox{\columnwidth}{!}{$
H_{01}\oplus H_{10}
\;\to\;
\left(
\begin{smallmatrix}
H_{01}&0&V_{01,\alpha}\\
0&H_{10}&V_{10,\alpha}\\
V_{\alpha,01}&V_{\alpha,10}&H_{\alpha}
\end{smallmatrix}
\right)
\;\to\;
\left(
\begin{smallmatrix}
H_{01}&0&V_{01,00}&V_{01,11}\\
0&H_{10}&V_{10,00}&V_{10,11}\\
V_{00,01}&V_{00,10}&H_{00}&0\\
V_{11,01}&V_{11,10}&0&H_{11}
\end{smallmatrix}
\right)
\!\!,\ \alpha=00\ \text{or}\ 11
$}
\label{eq:block_connectivity}
\end{equation}

Shown in Eq.~\eqref{eq:block_connectivity} is how the block decomposition in the analytical limit of the ``hard'' junction is progressively destroyed as one or both intermediate junction sectors are activated. The Hilbert space fragmentation is not destroyed because of introduction of off-diagonal elements between $01$ and $10$ sectors because those elements do not exist for the single-flip Hamiltonian Eq.~\eqref{eq:composite_hamiltonian}. Rather the Hilbert space expands and single flip kinetic terms introduce $V_{\alpha,10/01}$ off-diagonal blocks polluting the direct sum structure that we had initially.

\section{Statistics and zero-modes}

In this section, at first, we will be concerned with only the $PXP$ and $d-PXP$ subchains intervened by a hard-junction. In that case, our projected Hamiltonian in the $01$ sector is $H=H_{l}^{L/2-1} \otimes|01\rangle\langle01|\otimes I_{r}+I_{l} \otimes|01\rangle\langle01|\otimes H_{r}^{L/2-1}$ which  means our eigenstates are of the form $|\psi_{l}\rangle  |01\rangle  |\psi_{r}\rangle$ where the individual $PXP$ and $d-PXP$ subchains are of length $L/2-1$. In the $10$ sector, the projected Hamiltonian is $H=H_{l}^{L/2-2} \otimes|0101\rangle\langle0101|\otimes I_{r}+I_{l} \otimes|0101\rangle\langle0101|\otimes H_{r}^{L/2-2}$. The sites adjacent to the junction are frozen to $0_{l}$ and $1_{r}$ and hence the effective $PXP$/ $d-PXP$ subchain lengths are $L/2-2$. The eigenstate in the $10$ sector can be written as $|\psi_{l}\rangle  |0101\rangle  |\psi_{r}\rangle$. The  spatial reflection symmetry $R_{l,r}$ maps site  $i \rightarrow L_{l,r}-i+1$ about the centers of the individual left and right sub- $PXP$ chains. The reflection operators $R_{l}\otimes |01\rangle\langle01|\otimes I_{r}$ and $I_{l}\otimes |01\rangle\langle01|\otimes R_{r}$ commute with the Hamiltonian $H_{01}$. 
\begin{figure}[h!]    \centering
    \includegraphics[width=0.45\linewidth,height=5.8cm]{ 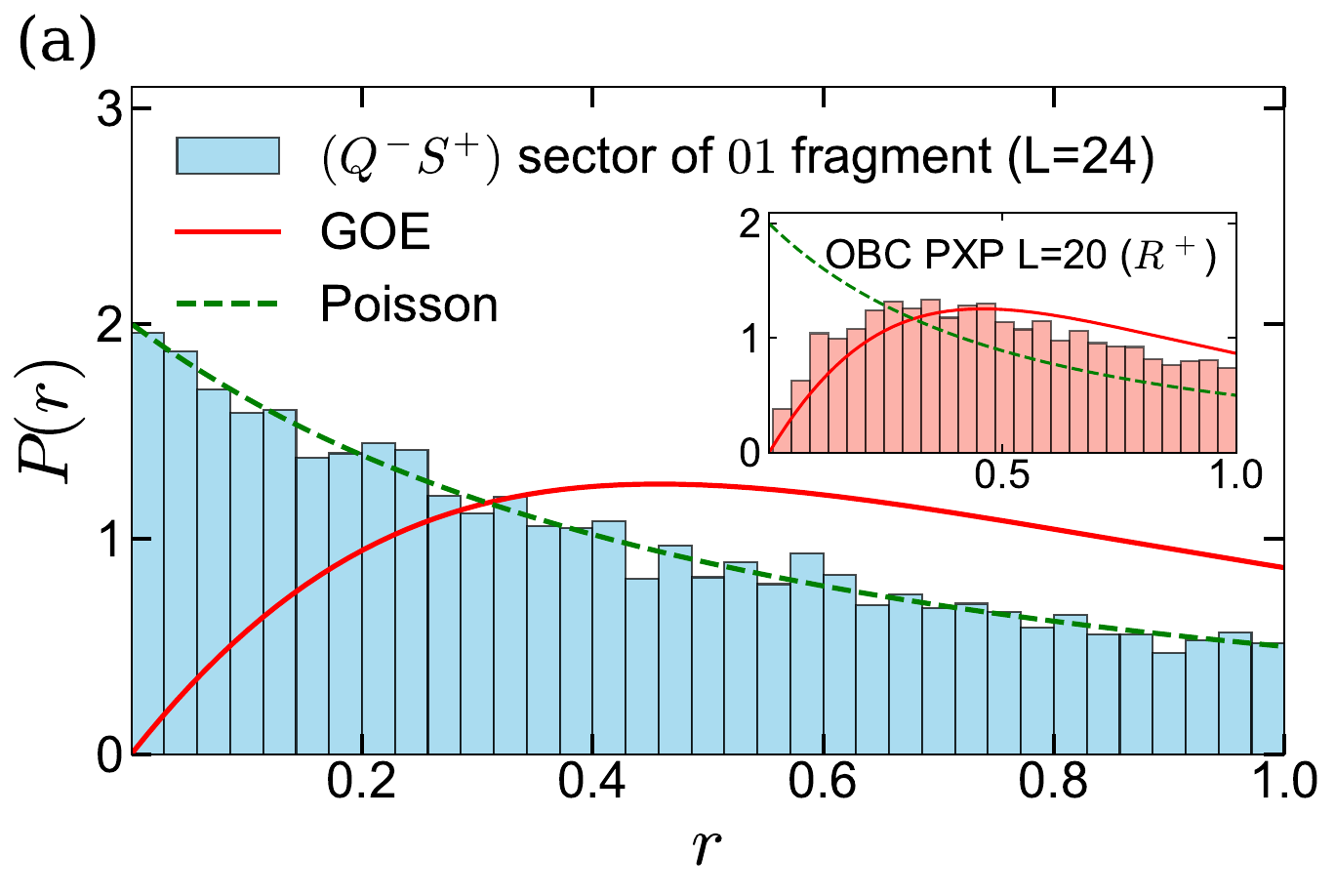}
    \includegraphics[width=0.53\linewidth,height=5.8cm]{ 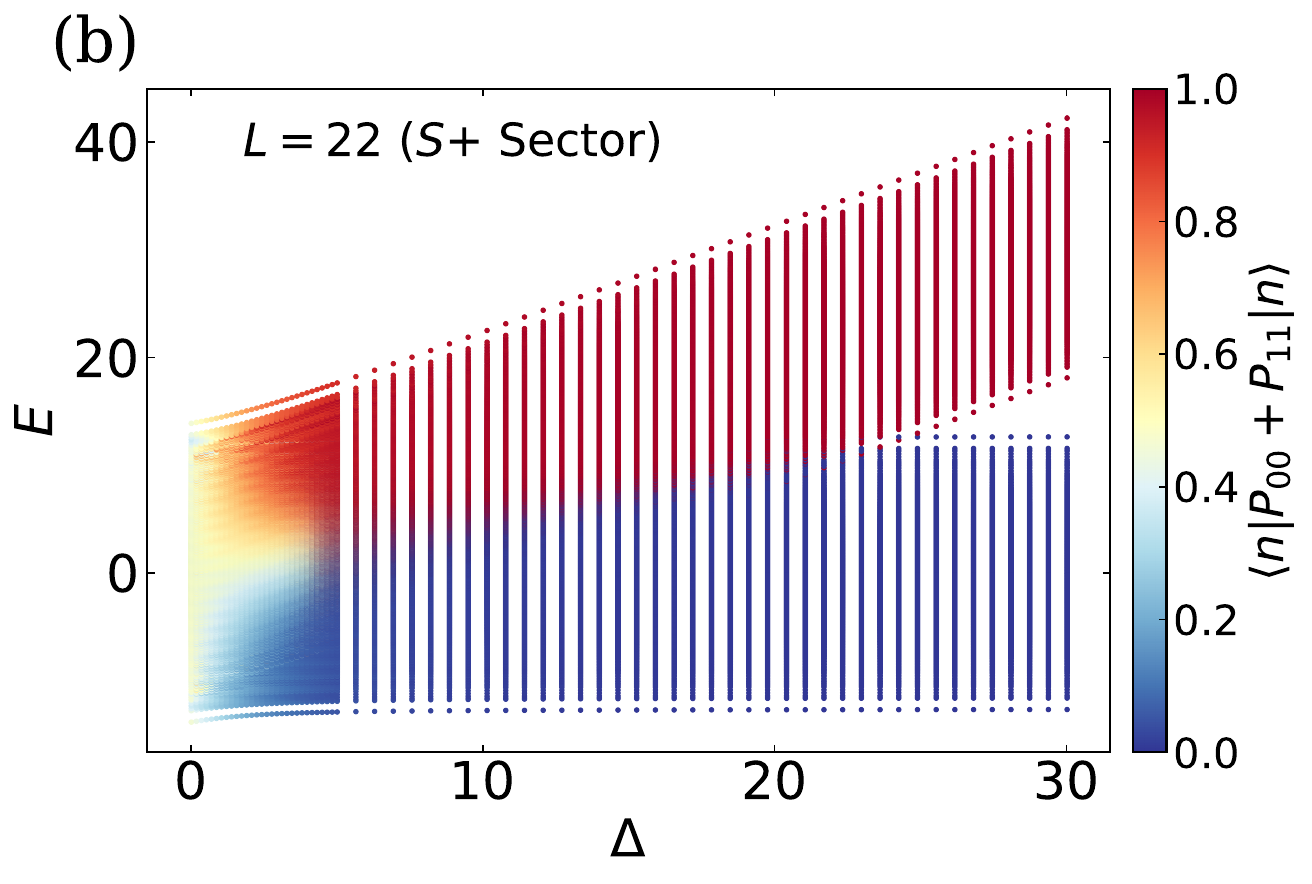}
    \includegraphics[width=0.53\linewidth,height=5.5cm]{ 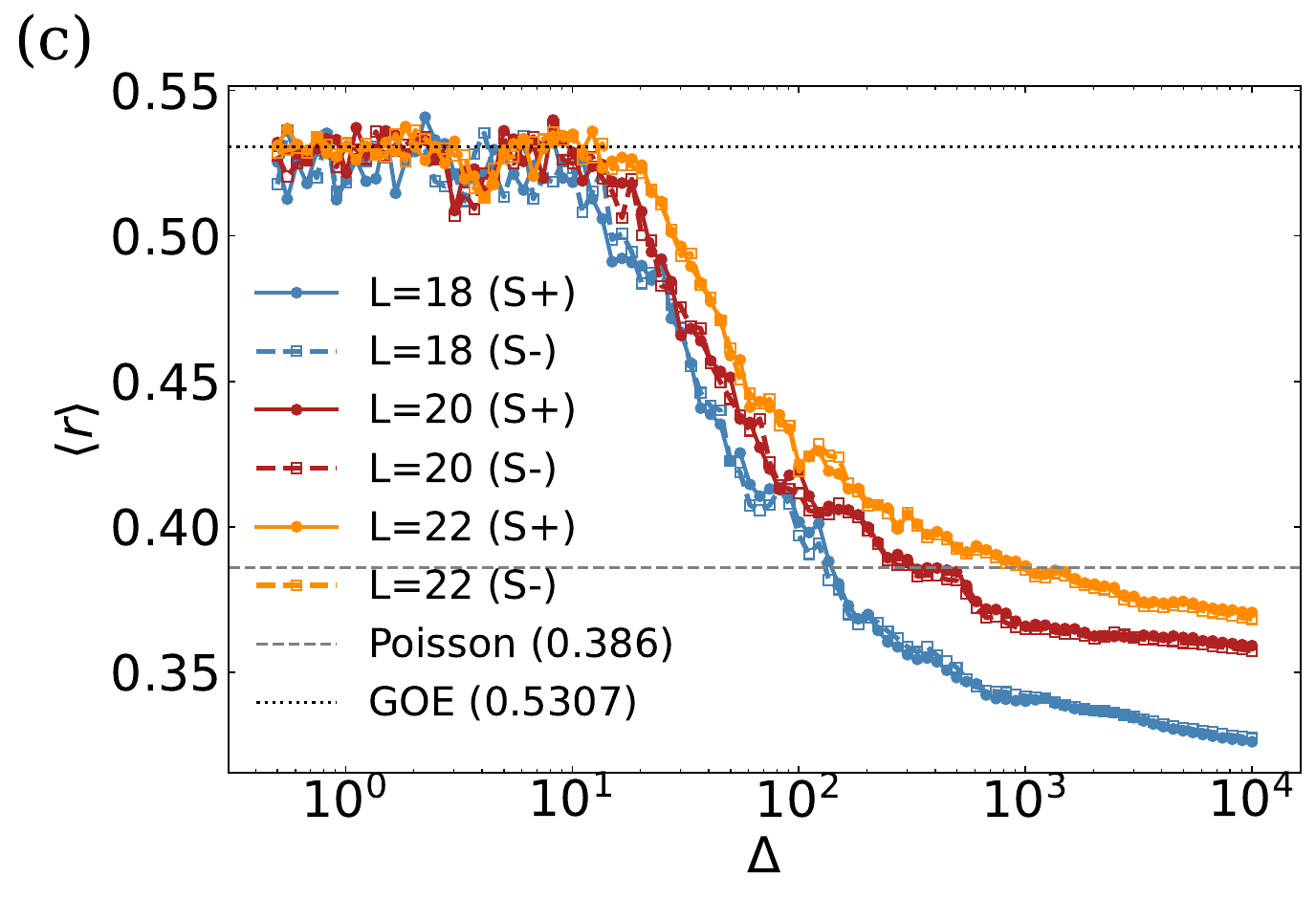}
    \caption{Panel $(a)$ shows that the level statistics of a $PXP-d$-$PXP$ chain of length $L=24$, with a hard-junction, in the $Q=-1, S=+1$ sector of the $01$ fragment is Poissonian in contrast with the $WD$ statistics of a pure $PXP$ chain of length $L=20$ in the $R=+1$ sector, which is shown in the inset. Panel $(b)$ shows the energy eigenvalues plotted against $\Delta_{00}=\Delta_{11}=\Delta$, in the $S=+1$ sector, for the soft-junction case where the colour coding of the data points has been used to show the expectation value of the $00$ and $11$ projectors at the junction, $\langle n|P_{00}+P_{11}|n\rangle$ in the corresponding energy eigenstate $|n\rangle$. This shows a gradual separation of the two groups of eigenstates, one with maximal weight on Fock states with $00$ and $11$ at the junction and the other with maximal weight on Fock states with $01$ and $10$ at the junction, with $\Delta$. Panel $(c)$ shows that the average value of the level spacing ratio $\langle r\rangle$ in the $S=\pm 1$ symmetry-resolved sectors for $L=18,20,22$, goes gradually from $WD$ to Poissonian with increasing $\Delta$, apart from an intermediate non-monotonicity.}
    {\label{fig2}}
\end{figure}
Therefore, this fragment can be block diagonalized in the $(R_{l},R_{r})=(+,+),(+,-),(-,+)$ and $(-,-)$ sectors. There is another internal symmetry  which is spatial reflection $R$ about the center of the entire chain combined with a bit-flip/ spin-flip operation $X$ which is given by $\prod_{i}\sigma_{x}^{i}$. We call this composite transformation $S=XR$ which commutes with the Hamiltonian and has eigenvalues $+1$ and $-1$. We can further decompose the $(+,+)$ and $(-,-)$ sectors into $S=\pm1$ symmetry sectors by taking suitable linear combination of the basis states in these blocks. When $S$ operates on the basis states of the $(+,-)$ sector, it maps to the basis states of the $(-,+)$ sector. Therefore, the $(+,-)$ and $(-,+)$ become connected under the symmetry operation $S$. Hence if we define, $Q=R_{l}R_{r}$, then the $Q=-1$ sector can be further resolved into two symmetry sectors $S=\pm 1$. The same arguments hold for the $10$ fragment. In addition, like the $PXP$ chain, there is chiral symmetry in the $PXP-d$-$PXP$ model where $C=\prod_{i=1}^{L_{l}}(-1)^{n_{l}^{i}}\times\prod_{i=L_{l}+1}^{L_{l}+L_{r}}(-1)^{n_{r}^{i}}$. The excitation on the left half of the chain is an up-spin i.e., the boson occupancy is  $1$ and on the right half of the chain is a down-spin i.e., the boson occupancy is  $0$. The operator $n^{i}_{l}$ measures whether the $i-$ th site is in the excited state and is equal to $1$ in that case, otherwise it is zero whereas  $n^{i}_{r}$ is the complement number operator defined on the right side where the excitations are holes. 
The chirality operator anti-commute with the Hamiltonian and ensures $E\leftrightarrow -E$ symmetry in the model. The $R_{l}\otimes |j\rangle\langle j|\otimes I_{r},I_{l}\otimes |j\rangle\langle j|\otimes R_{r}$ and $S$ operators commute with the $C$ operator.  If we make a basis transformation to go to a specific $S$ symmetry sector, then the level spacing statistics turns out to be Poissonian~\cite{ETH3,RMT} in contrast to the Wigner-Dyson (WD)~\cite{ETH3, RMT} statistics shown by each sub $PXP$ chain.  This is shown in figure~\ref{fig2}(a). When we talk about the chaotic spectrum of the PXP subchains then we are looking at a particular symmetry sector which happens to be $R=+ 1$ or $R=-1$, as shown in the inset of figure~\ref{fig2}(a). Here, we observe that the simultaneous eigenstates of $H$ and $S$ in the $S=\pm1$ sector of the $01$ fragment in the  respective $(+1, +1)$, $(-1, -1)$ or $Q=-1$ sectors can be written as,
\begin{equation}
\begin{aligned}
&\sqrt{2}|\psi_{S=\pm1}^{++}\rangle=|\psi_{l}^{+1}\rangle |01\rangle|\psi_{r}^{+1}\rangle\pm|\tilde{\psi}_{r}^{+1}\rangle |01\rangle|\tilde{\psi}_{l}^{+1}\rangle\\ 
&\sqrt{2}|\psi_{S=\pm1}^{--}\rangle=|\psi_{l}^{-1}\rangle|01\rangle |\psi_{r}^{-1}\rangle\pm|\tilde{\psi}_{r}^{-1}\rangle  |01\rangle  |\tilde{\psi}_{l}^{-1}\rangle\\  &\sqrt{2}|\psi_{S=\pm 1 }^{Q=-1}\rangle=|\psi_{l}^{\pm 1}\rangle  |01\rangle  |\psi_{r}^{\mp 1}\rangle\pm|\tilde{\psi}_{r}^{\mp 1}\rangle  |01\rangle  |\tilde{\psi}_{l}^{\pm1}\rangle.\\ 
\end{aligned}
\label{eq:simulSH}
\end{equation}

Here, $|\tilde{\psi}_{l,r}\rangle=X |\psi_{l,r}\rangle$. Since,

\begin{equation*}
\begin{aligned}
&\langle \psi_{r}^{+/-}| H_{d-PXP}^{L/2-1}|\psi_{r}^{+/-}\rangle=\langle \tilde{\psi}_{r}^{+/-}| H_{PXP}^{L/2-1}|\tilde{\psi}_{r}^{+/-}\rangle \\ 
&\langle \psi_{l}^{+/-}| H_{PXP}^{L/2-1}|\psi_{l}^{+/-}\rangle=\langle \tilde{\psi}_{l}^{+/-}| H_{d-PXP}^{L/2-1}|\tilde{\psi}_{l}^{+/-}\rangle,\\
\end{aligned}
\end{equation*}

the energy eigenvalues of the two parts in the symmetric/anti-symmetric combination in Eq.~\eqref{eq:simulSH} are the same. Therefore, the energy eigenvalues of these eigenstates are $E_{l}^{\pm}+E_{r}^{\pm}$ or $E_{l}^{\pm}+E_{r}^{\mp}$, where $E_{l,r}^{\pm}$ are the eigen-energies of the left $PXP$ chain and the right $d-PXP$ chain in the respective $R_{l,r}$ sectors. Therefore, the spectrum is a pairwise sum of two dynamically independent subchain spectra which according to random matrix theory (RMT)~\cite{ETH3,RMT} should not show level repulsion due to interpenetration of the individual Wigner-Dyson (WD) spectra.

Now, let us study what happens to the statistics for the soft junction, where for simplicity we have considered $\Delta_{00}=\Delta_{11}=\Delta$. As we progressively increase $\Delta$, the average value of the  level repulsion ratio $\langle r\rangle$ in the $S=\pm1$ symmetry-resolved sector gradually goes from $\langle r\rangle=0.53$ (WD) to  $\langle r\rangle=0.38$ (Poissonian) which is the asymptotic limit at $\Delta \rightarrow \infty$. In the panel $(c)$ of the figure~\ref{fig2}, we show the $\langle r\rangle$ versus $\Delta$ plot for $L=18,20,22$. However, there is a non-monotonicity in the curve which requires further explanation. We see that although initially for weak $\Delta>0$ the statistics is WD, as $\Delta$ is increased higher there is a visible dip in the statistics which rises again and then decays monotonically with increasing $\Delta$. This behaviour can be understood with the help of panel $(b)$ of figure~\ref{fig2} where the energy eigenvalues in the symmetry-resolved sector are plotted on the y-axis and $\Delta$ on the x-axis. The energy eigenvalues are colour marked by $\langle n|(|00\rangle\langle 00|+|11\rangle\langle11|)|n\rangle$  which is the  expectation value of the $00$ and $11$ projectors at the  junction sites between the corresponding eigenstate  $|n\rangle$ of the system in the said symmetry sector. Colours in the lower frequency of the visible spectrum (towards red) signify higher expectation values. Initially we see that there is no separation between the two classes of eigenvectors, that is the ones which have majority weight on the Fock states with $00$ and $11$ at the junction sites  and the ones which have majority weight on Fock states with  $01$ and $10$ at the junction sites. Therefore, they form a fully chaotic dense spectrum with level repulsion because the mean many-body level spacing $\delta$ is much smaller than the hybridization term $\epsilon=1$ which creates avoided crossings in the spectrum.  With increasing $\Delta$, the former group tries to separate from the latter group, increasing $\delta$ while suppressing $\epsilon/\delta$ and hence the hybridization, which creates a dip in $\langle r \rangle$. With further increasing $\Delta$,  the center of the $00$ and $11$ group try to move upwards while  the center of the $10$ and $01$ group remains the same. The tail of the prior group encounters few resonances as it moves past the many-body energy spectrum of the latter group, resulting again in avoided crossings. This results in the increase in $\langle r\rangle$ from  the dip value. As the centers of the two groups of many-body spectra separate without overlap, $\langle r\rangle$ decays towards the Poissonian value. The effective hybridization parameter in the large $\Delta$ perturbative limit is $\epsilon^{2}/\Delta \ll \delta$ which progressively makes the two groups independent. The derivation of the effective Hamiltonian from Schrieffer-Wolff transformation~\cite{Bravyi2011} is detailed in the Appendix A. As $\Delta \rightarrow \infty$, the former group gets decoupled from the latter group and the many-body spectrum is made up of solely $01$ and $10$ junction states which do not talk to each other. This direct sum Hamiltonian gives us a Poisson distribution, but this does not eliminate the possibility that the $01$ and $10$ sectors may have WD type statistics individually. This possibility is discarded as shown in panel (a) of the figure, where in the sector where all commuting symmetries are resolved, the statistics is still Poissonian which is because of the pairwise sum of the chaotic spectra of the two disjoint subchains intervened by the hard-junction. It is to be noted that $S$ is the relevant commuting symmetry for $\Delta > 0$ unlike the special points $\Delta=0$ and $\Delta \rightarrow \infty$ limit where we can have additional symmetries. It is to be noted that the special points \(\Delta=0\) and \(\Delta\rightarrow\infty\) remain Poissonian, whereas for any fixed finite \(0<\Delta<\infty\) the statistics flows toward WD with increasing system size. This is because the many-body energy spacing goes like $\delta(L)\sim e^{-L}$ while the bandwidth of both the groups we discussed earlier scales with $L$, leading to chaotic behaviour. Our finite-size data in figure~\ref{fig2}(c) are consistent with this expectation. We also observe in figure~\ref{fig2}(c) that for large $\Delta$, $\langle r\rangle$ goes below the Poissonian value showing sub-Poissonian statistics. But this is a finite-size cross over and $\langle r\rangle$ increseases gradually to $\langle r\rangle=0.38$ as the system size is progressively made larger.

In the $\Delta \rightarrow \infty$ limit of this model, the zero-modes are of two kinds: accidental ones which are formed from $(E,-E)$ pairs taken from the left and right chain spectra and product ones which come from $(0,0)$ modes of the individual subchains. Since $\{C,H\}=0$, $[H,S]=0$  and $[C, S]=0$, the sector-wise index theorem~\cite{Index, Index1} dictates that the total number of zero-modes in each $S$ sector is lower bounded by the imbalance/ Witten index, $|N_{S}^{+}-N_{S}^{-}|$, where $N_{s}^{\pm}$ is the total number of basis states in the $S$ sector with $\pm 1$ chirality. The number of accidental zero-modes will be equal to $\sum_{E\ne 0}g(E)g(-E)$ where the dimension of the degenerate manifold corresponding to energy $E\ne 0$ is $g(E)$ which is equal to $g(-E)$ due to the $E\leftrightarrow-E$ symmetry in the spectrum. The number of product modes is equal to $N^{PXP}_{zero} \times N^{PXP}_{zero}$. It is to be noted that the energy spectra of the $PXP$ and $d-PXP$ subchains of equal lengths are exactly identical. 

\begin{figure}[h!]
    \centering
    \includegraphics[width=0.48\linewidth,height=5.5cm]{ 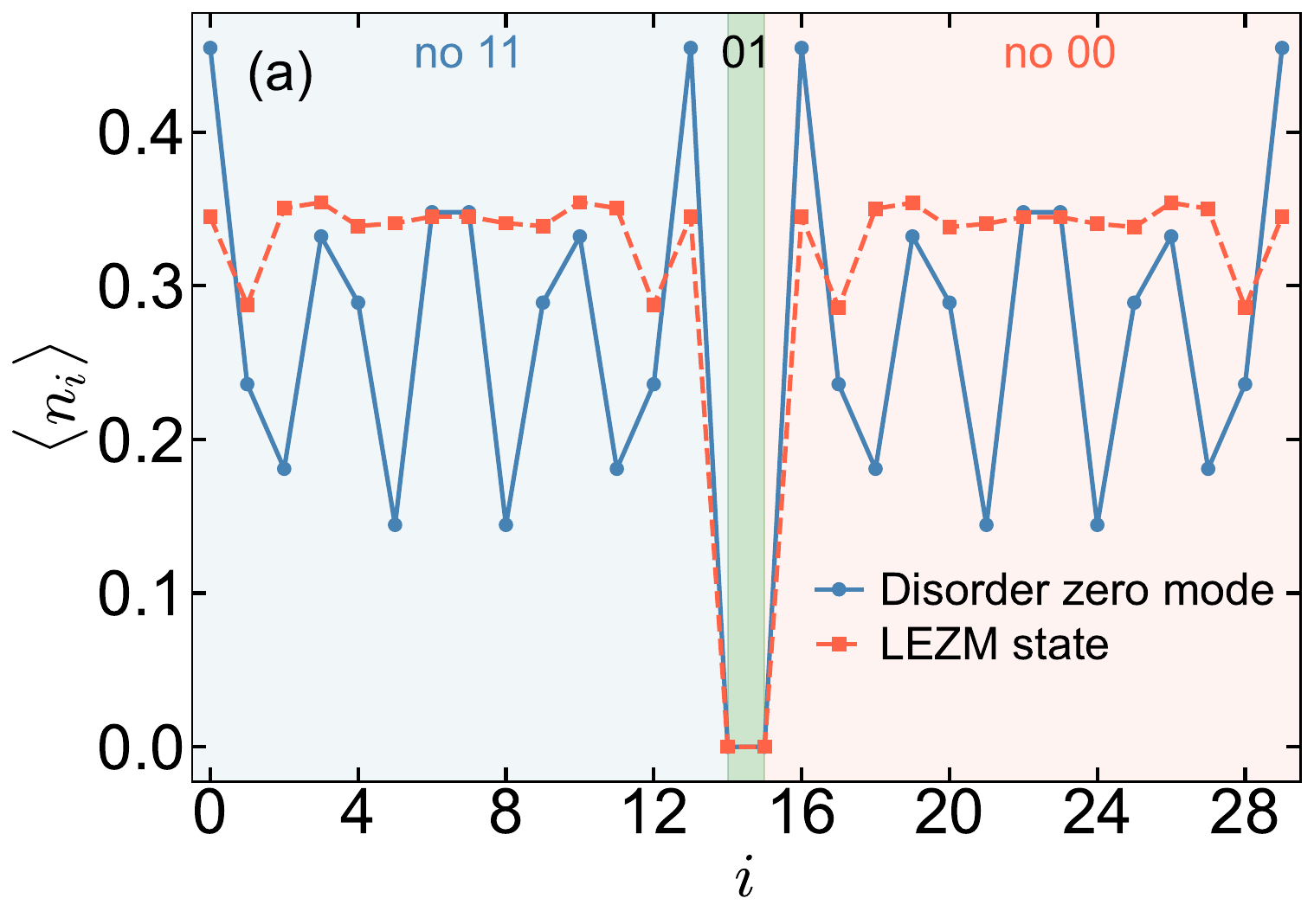}    
    \hspace{0.01\linewidth}
    \includegraphics[width=0.48\linewidth,height=5.5cm]{ 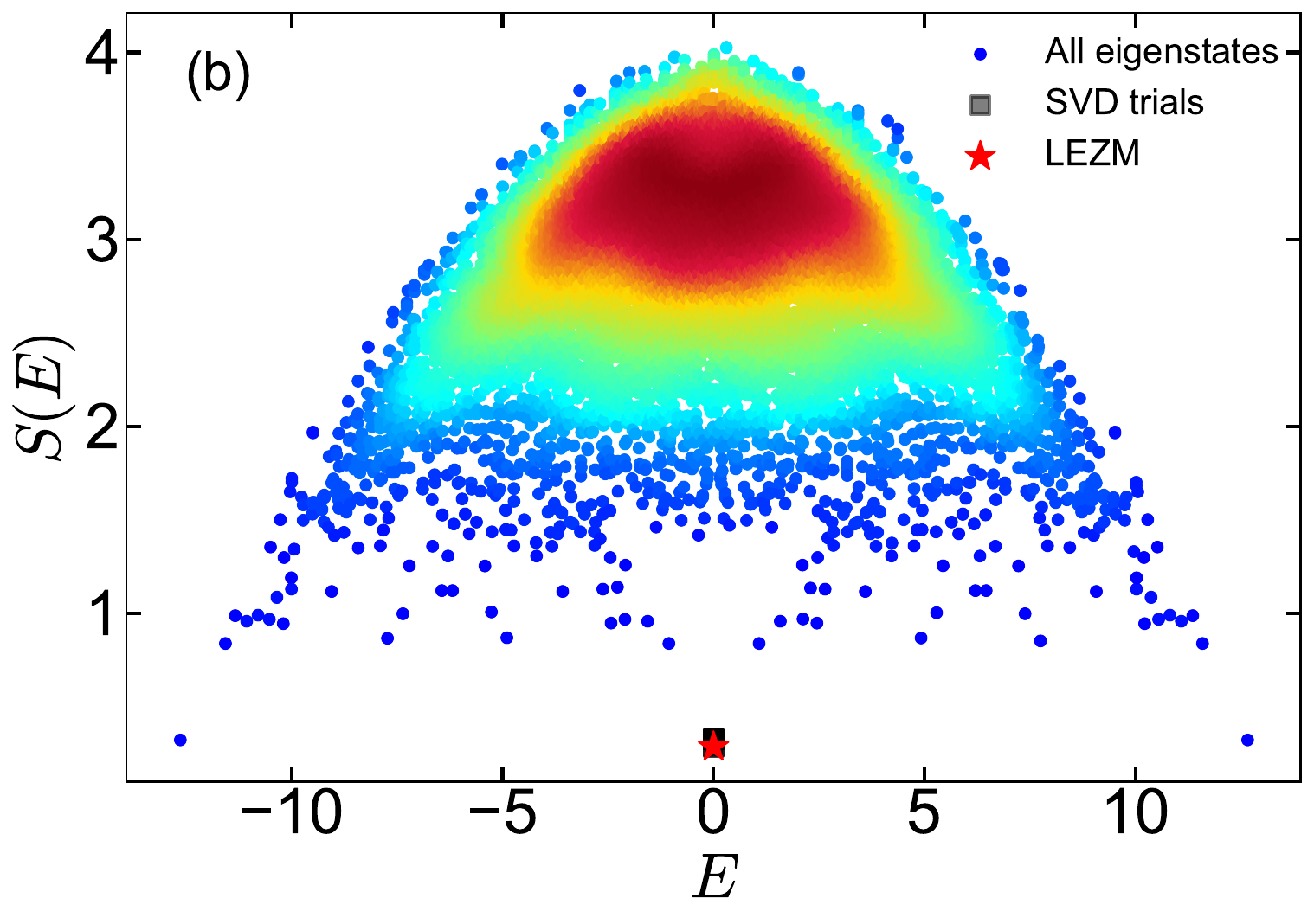}
    \includegraphics[width=\linewidth,height=4cm]{ 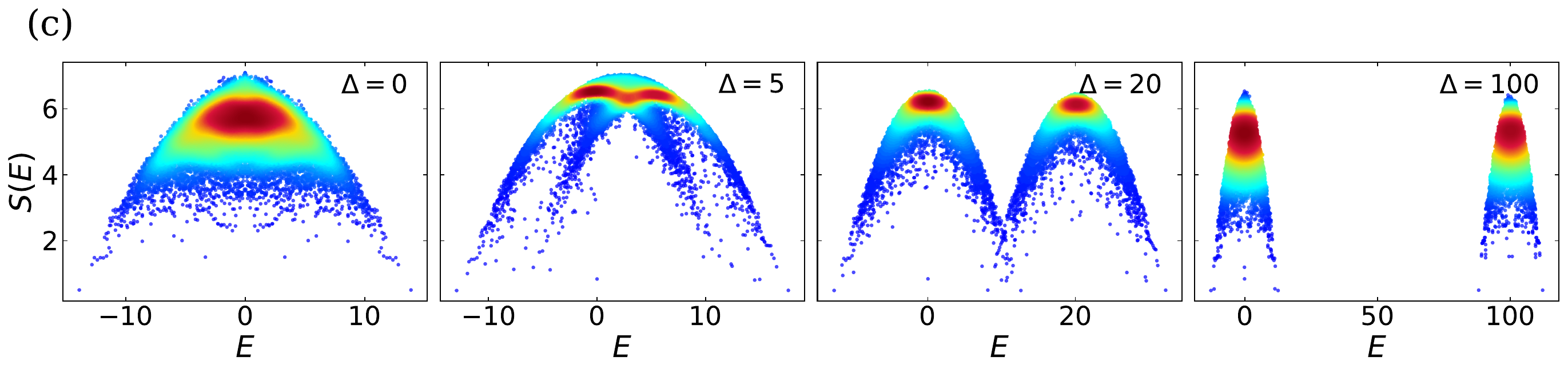}
    \caption{Panel (a) shows the density profile $\langle n_l^{i} \rangle$ on the left half of the chain and $\langle n_r^{i} \rangle$ on the right half of the chain for the tensor product state $|0_l\rangle|01\rangle|0_r\rangle$ in case of a hard-junction. The blue curve corresponds to the chirally protected zero-mode, obtained by introducing a random $PXP$ disorder into the individual subchains ($L_{\text{sub}}=14$) with amplitudes $\epsilon_j$ drawn from a uniform distribution $\epsilon_j \in [0, 10^{-5}]$. The red curve shows the tensor product of the least entangled zero-modes (LEZM) of the respective subchains, obtained via an entanglement minimization algorithm in the pristine limit. Panel (b) shows the bipartite Von Neumann entanglement entropy $S_E$ for the $01$ sector of the composite hard-junction PXP-$d$-PXP chain ($L=22$), utilizing a two-cut bipartition at $x=5$ and $x=15$ bonds. The LEZM for this composite model, isolated via the SVD minimization technique, is additionally highlighted. Panel (c) shows the bipartite entanglement entropy $S(E)$ in the $S=+1$ sector for  two cuts at $x=6$ and $x=17$ bonds, for different values of $\Delta_{00}=\Delta_{11}=\Delta$ in the soft-junction case for $L=22$. It gradually goes from a unimodal structure at $\Delta=0$ to a bimodal structure, in the $S(E)$ vs $E$ plane, with increasing separation between the peaks, as $\Delta$ is increased. As $\Delta \rightarrow \infty$, the peak corresponding to the eigenstates with weight on the Fock states with $00$ and $11$ at the junction moves to infinity and the only peak is the one which has eigenstates with weight on Fock states with $01$ and $10$ at the junction sites which is shown in panel $(b)$ (the hard-junction case).  }
    {\label{fig3}}
\end{figure}

If the Hilbert space dimension of the individual subchains are odd, then the $|\mathrm{Tr}(C)|=1$ for each subchain, which indicates the presence of at least one chirally protected mode. We observe that the zero-mode subspace in each $R$ sector is chirally locked. We also observe that the zero-mode dimension of the $R$ sectors differ by one and hence there is one protected mode which lives in the larger sector. To pin down the chirally protected mode, we introduce a chirality-preserving disorder, $\sum_{i}\epsilon_{i}P^{i-1}_{l,r}\sigma_{X}^{i}P^{i+1}_{l,r}$, where $\epsilon_{i}\ll 1$ is a weak random disorder. As a consequence of this, only the protected mode remains pinned to zero eigenvalue while the rest of the zero-modes hybridize. We extract these modes for a $PXP$ chain as well as a $d$-$PXP$ chain of equal lengths. Then the required chirally protected zero-mode of the composite model is $|0_{l}\rangle|01\rangle|0_{r}\rangle$ whose density profile is shown in figure~\ref{fig3}(a). The profile clearly shows that there are four dominant peaks with a frozen junction in the middle. The junction broadens if we go to the $10$ sector rather than the $01$ sector. These edge modes remain robust under any disorder which preserves chiral symmetry and therefore can be used for storing quantum information as they do not easily hybridize with the bulk states. However, they do have high entanglement entropy. This discussion holds for the hard constraint at the junction. In the soft-junction case, for $\Delta_{11}=-\Delta_{00}$ and $\Delta_{11,00}\ne0$, the usual chirality operator \(C\) no longer anticommutes with the Hamiltonian; instead, the combined operator \(\Gamma=CXR\) provides the chiral symmetry, \(\{\Gamma,H\}=0\), which imposes  $E \leftrightarrow -E$  symmetry in the spectrum. We introduce a term $\sum_{i,j}\epsilon_{ij}(|\mathcal{F}_{i}^{L/2-1}\rangle\langle\mathcal{F}_{i}^{L/2-1}|\otimes|00\rangle\langle00|\otimes|1\rangle\langle1|\otimes|\mathcal{F}_{j}^{L/2-2}\rangle\langle\mathcal{F}_{j}^{L/2-2}|-|\tilde{\mathcal{F}}_{j}^{L/2-2}\rangle\langle\tilde{\mathcal{F}}_{j}^{L/2-2}|\otimes|0\rangle\langle0|\otimes|11\rangle\\\langle11|\otimes|\tilde{\mathcal{F}}_{i}^{L/2-1}\rangle\langle\tilde{\mathcal{F}}_{i}^{L/2-1}|)$, where $\epsilon_{ij}\ll 1$ is a weak random disorder, in the Hamiltonian which anticommutes with $\Gamma$ and observe that all the zero modes here are protected against this chiral preserving disorder. These zero modes generally have usual peaked structure at the boundaries and near the junction. For $\Delta_{00}=\Delta_{11}=\Delta$, there is no chiral operator that anticommutes with the Hamiltonian and consequently the zero-modes are not protected. For $\Delta=0$, the chain can be factorized into two decoupled $PXP$ and $d-PXP$ chains of length $L/2$. Therefore, like the hard junction, here we also have accidental and product zero modes. Here, the chirality operator $C$ protects one zero mode for Hilbert space dimensions of $PXP/d-PXP$ sub-chains which are odd.

 We also execute an entanglement minimization algorithm~\cite{SVD,SVD1} using  singular value decomposition ($SVD$) in the zero-mode space of each subchain and obtain the tensor product eigenstate from which we plot the density profile, in case of the hard junction. This zero-mode state has anomalously low entanglement entropy in contrast to the earlier chirally protected state and is athermal. This is also shown in c. It is important to mention that the peaked structures in the density profile are not unique to the zero-mode subspace but appear generically throughout the spectrum as a consequence of the geometric cleaving. In figure~\ref{fig3}(b), we show the Von-Neumann entanglement entropy of the system (in the $01$ sector) with two cuts, one in the bulk of the left chain and the other in the bulk of the right chain. The extreme two subparts are treated as one subsystem whereas the middle portion harbouring the junction is treated as the other subsystem for calculating the bipartite entanglement entropy. The striking feature of this plot is the multiplicity of the sub-volume scar states with nonzero energies. Because of the tensor product of the states in the scar manifold, we find, the multiplicity of these anomalous states increases as compared to a homogeneous $PXP$ chain of equal length. Further, we have also marked the entanglement entropy of the zero-mode state obtained from entanglement minimization algorithm using the $SVD$ technique in the composite model. In  case of the soft-junction, in the panel (c) of the figure~\ref{fig3} we show the two-cut bipartite entanglement entropy as $\Delta$ is increased. We see that at low values of $\Delta$ a unimodal structure exists because the two groups of eigenstates, one dominated by $00,11$ junction Fock states and the other dominated by $01,10$ junction Fock states, were overlapping with each other in eigenenergies. Therefore, they constitute a quasi-continuous spectrum. However, as we start increasing $\Delta$ the unimodal structure starts separating to give rise to a bimodal structure, in the $S(E)$ vs $E$ plane, with two prominent peaks, solely due to separation in the energy scales. The peak corresponding to the former group moves right on the energy scale as $\Delta$ is increased while the peak associated with the latter group is pivoted. Finally, as $\Delta \rightarrow \infty$, the former peak moves to infinity, which tells us that our relevant Hilbert space consists only of Fock states with $01/10$ at the junction and we retrieve the unimodal structure of panel (b) of the figure~\ref{fig3}.

\section{Dynamical signatures}

\begin{figure}[ht!]
    \centering
    \includegraphics[width=\linewidth,height=10.0cm]{ 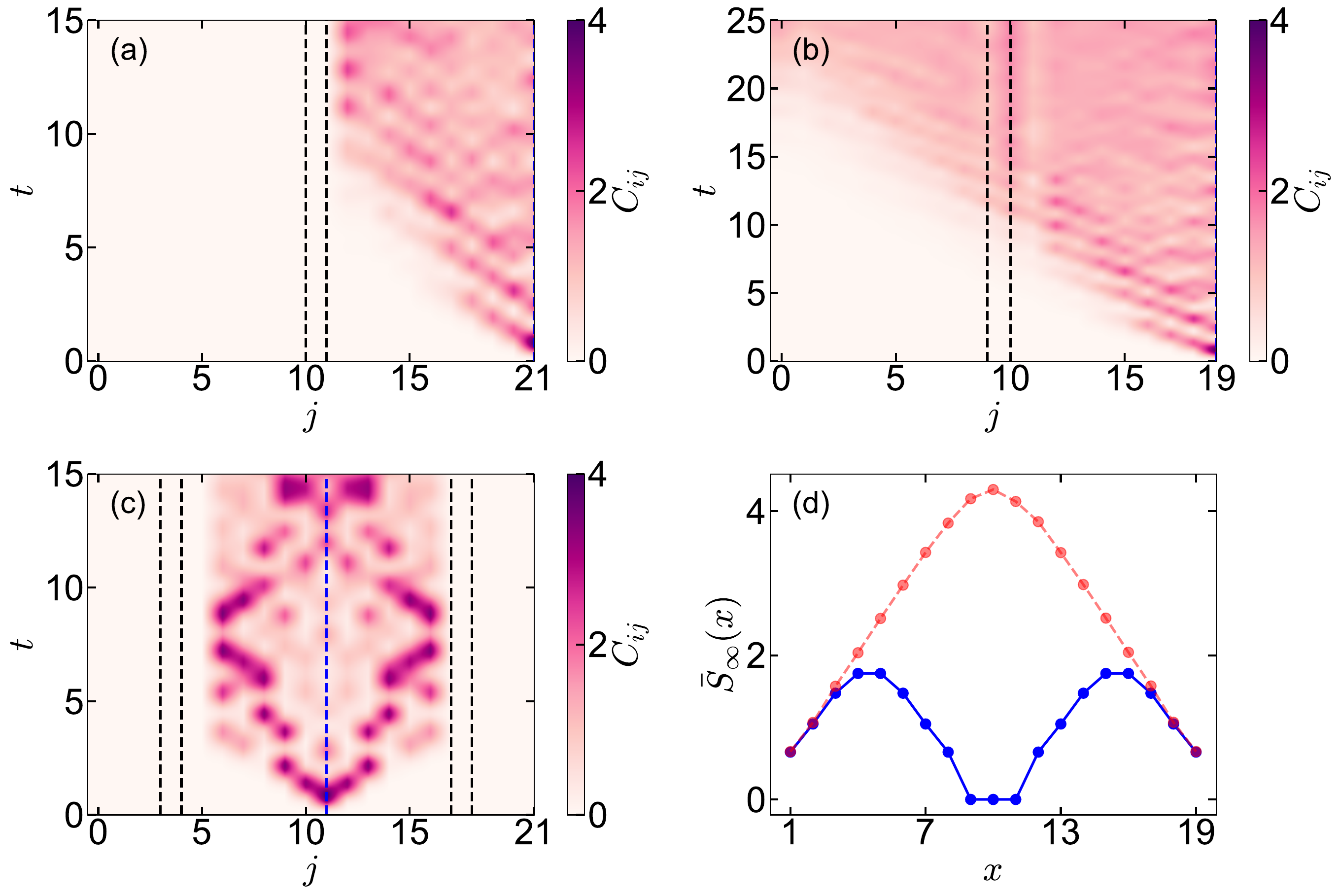}
     \caption{Panel $(a)$ shows that the $OTOC$, $C_{ij}$ for a hard-junction $PXP-d$-$PXP$ chain of length, $L=22$ in the initial state $|00..00|01|11..10\rangle$ with site $i=21$, is confined to $j>11$ under unitary time evolution. In panel $(b)$ we show the same when the $00$ constraint is lifted for $L=20$. In panel $(c)$, we show the caging effect of quantum information between two junctions in the initial state $|000|10|1010110_{i}11010|10|000\rangle$($L=22$). Panel $(d)$ shows  the late-time average of the entanglement entropy, $\bar{S}_{\infty}(x)$ as a function of cuts along the chain where cuts, $x$ are made on the different bonds, for a $PXP-d$-$PXP$ chain in its vacuum state, $|00..0|01|1..11\rangle$ $(L=20)$. Here, the blue curve shows the case when  both $00$ and $11$ are not allowed at the junction and the red curve shows the same when the junction becomes permeable and allows only $00$.} 
    \label{fig4}
\end{figure}

\begin{figure}[ht!]
    \centering
    \includegraphics[width=\linewidth,height=5.8cm]{ 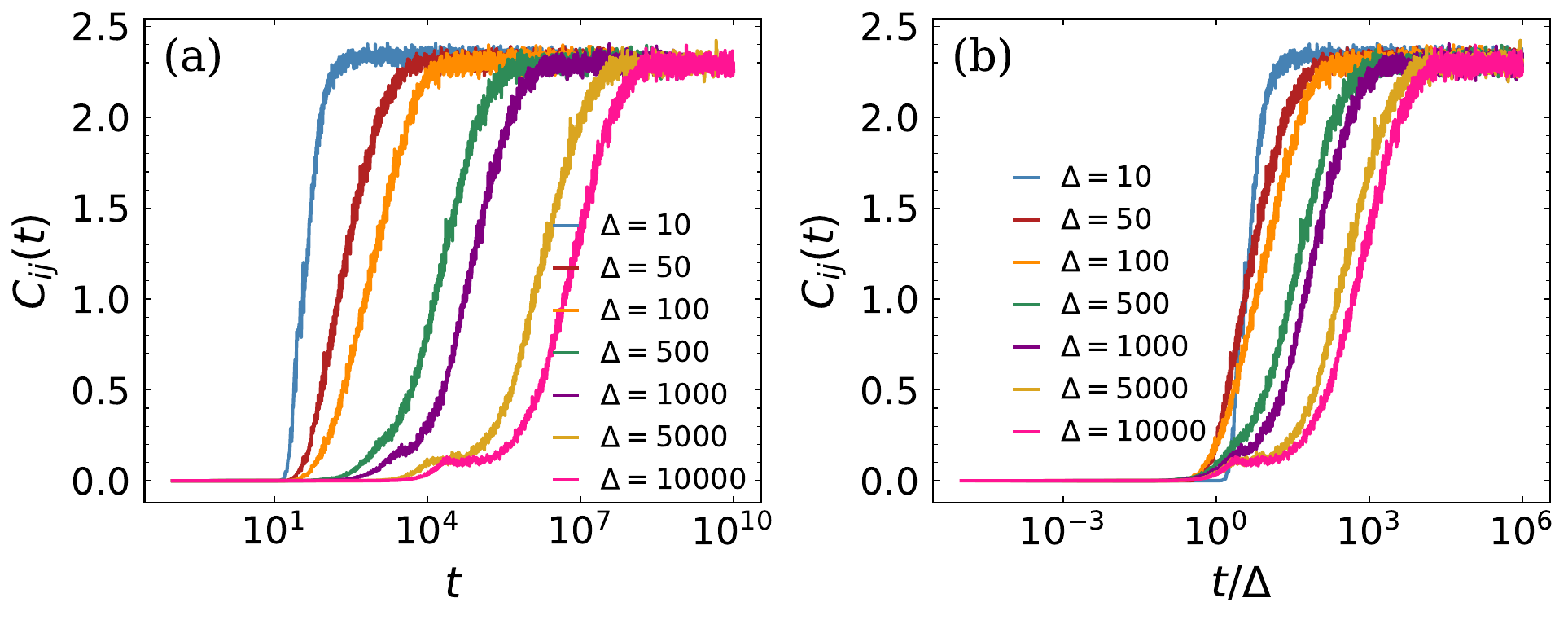}
    \includegraphics[width=\linewidth,height=5.8cm]{ 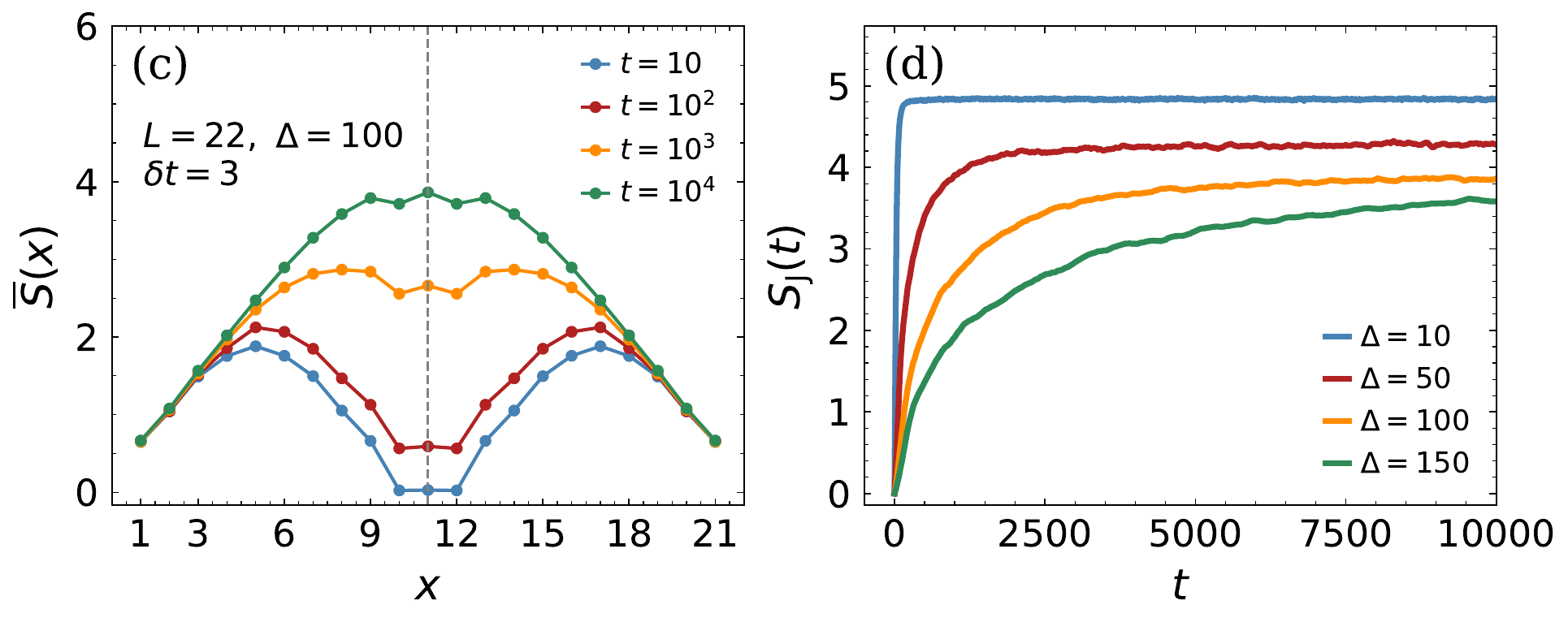}
    \caption{Panel $(a)$ shows how the $OTOC$, $C_{ij}(t)$ varies as a function of time $t$ for a $PXP-d$-$PXP$ chain with a soft-junction for different values of $\Delta_{00}=\Delta_{11}=\Delta$ for $L=20$ with the source site $i=1$ and target site $j=20$ for the vacuum state. The larger the value of $\Delta$ larger is the onset time $t^{*}_{o}$ for the information to leak from site $i$ to $j$. For $t>t^{*}_{o}$, the information received at site $j$ gradually increases to reach a saturation value at $t^{*}_{\Delta}$, beyond which it does not increase further. In panel $(b)$ we show that the onset time $t^{*}_{o}$ scales as $\Delta$ independent, within numerical precision, of the initial state we start the dynamics from. Therefore, the junction behaves practically as a hard-junction separating the dynamics of the subchains up to a timescale $t^{*}_{o}\sim \Delta$. Panel $(c)$ shows how the entanglement entropy, $\bar{S}(x)$ varies as a function of the bond cut $x$ for $L=22,\Delta=100$, at different times $t$, averaged over a small window $[t-\delta t,t]$ where $\delta t=3$, starting from the vacuum state at $t=0$. Initially for $t<t^{*}_{o}$, the entanglement entropy across the junction cut is indistinguishable within resolution from an ideal hard-junction in this time range and we have a bimodal structure. Beyond this, for $t^{*}_{o}<t<t^{*}_{\Delta}$  the junction entanglement entropy rises gradually to its saturationvalue and the structure becomes unimodal as is expected for a chain which has undergone bulk thermalization. Panel $(d)$ shows the variation of junction entanglement entropy $S_{J}(t)$ with $t$ starting from the initial vacuum state for $L=22$. It shows greater the value of $\Delta$, greater is the prethermal window over which the gradual process of thermalization occurs. However, this timescale $t^{*}_{\Delta}$ is sensitive to the initial state we start from, although the trend that it increases with $\Delta$ is generic.   }
    {\label{fig5}}
\end{figure}

The most important consequence of Hilbert space fragmentation in this model with hard constraint at the junction, is the dynamic insulation of propagating information. Since the junction is frozen and there is no dynamical pathway between $01$ and $10$ sectors, a local operator initially supported on one fragment cannot spread into the other. The hard-junction therefore produces exact dynamical insulation of local operator information. In figure~\ref{fig4}(a), we give the plot of out-of-time-order correlator (OTOC)~\cite{OTOC1,OTOC2,OTOC3,OTOC4} which is defined as \[
C_{ij}(t)
= 2\left(1 - \operatorname{Re}\left(
\left\langle
\sigma_i^z(t)\,
\sigma_j^z\,
\sigma_i^z(t)\,
\sigma_j^z
\right\rangle\right)\right)
\] for the vacuum state of the system with a single site excitation i.e., $|00..00|01|11..10\rangle$. The $i-$ th site is fixed at the right most site and $j$ sites are varied through the chain. This shows how information is caged in one side of the chain in the presence of a rigid junction. As soon as we relax the constraints on the junction and allow either $11$ or $00$ at the junction sites , information leaks and this is how we can tune the constraint to allow/disallow information flow. This is shown in figure~\ref{fig4}(b). In figure~\ref{fig4}(c), we engineer two $10$ junctions and cage information in a shielded well, insulated on both sides by the rigid junction constraints. On the leftmost and rightmost sections $11$ is not allowed and in the middle region $00$ is not allowed.  The width of the regions with $C_{ij}=0$ is asymmetric because the frozen sites in case of the first $10$ junction are actually $0101$ unlike the second $10$ junction configuration. In both the hard junction cases, we have unitarily evolved the operators with respect to the total Hamiltonian which gives numerical evidence of Hilbert space fragmentation parametrized by frozen junction configurations, as is evident from the localization of the dynamics. Here, it is important to distinguish between two different yet intertwined aspects of information localization. For a finite density of hard junctions, the number of dynamically disconnected Krylov fragments grows exponentially with the system size \(L\). An initial state belonging to a particular Krylov fragment remains confined to that fragment under unitary evolution, thereby restricting the fraction of Fock space over which its information can spread. In this sense, Hilbert-space fragmentation leads to localization of information in Fock space. In the present model, however, there is an additional real-space localization mechanism. The hard junctions factorize the dynamics into a tensor sum of independent subchain Hamiltonians, so that spatially separated segments evolve unitarily and independently of one another. Information initially contained within a segment therefore cannot propagate across the neighboring hard junctions. As the junction density is increased, the intervening subsegments become shorter and the corresponding local Hilbert spaces available for scrambling become smaller. In the extreme limit in which only a single active site is isolated between neighboring junctions, its state undergoes coherent oscillations between the two local basis states. If the local unitary evolution of an isolated segment is known and coherently controllable, the initial-state information can, in principle, be retrieved at an arbitrary later time by reversing the local unitary evolution. Although this is true for any finite-size segment, smaller isolated segments confine the information to fewer sites and fewer accessible Fock states, making the dynamics simpler and the initial state easier to retrieve in practice. It is to be noted that, since the Krylov fragments are labeled by frozen junction configurations, the Fock space and real-space localizations are intrinsically coupled in this model.

Next, in fig~\ref{fig4}(d), we show the late-time average of the entanglement entropy $\bar{S}_{\infty}$ for different cuts in our chain. We see a $M$-shaped structure arising because the two halves on the left and right are like two thermal fluids separated by a knot at the junction. As we relax the constraints on the junction, either one or both, the two regions become dynamically connected and evolve toward a common thermalizing regime, as expected. So the junction acts as a knob for the control of information propagation. In the general soft-junction case, OTOC between two-sites $(i,j)$ on two sides of the junction for an initial Fock state, leaks from the source to the target site with an effective cross-junction coupling, inversely proportional to $\Delta$. The term $\Delta$ is akin to the height of the potential barrier between the two sides of the chain which controls the information tunneling amplitude. For large $\Delta$, the role of the tunneling amplitude is played by $J_{\rm eff}=-2\epsilon^{2}/\Delta$ ( Appendix A), the cross-junction effective coupling which for $\Delta \rightarrow \infty$, corresponding to an infinite barrier, goes to zero and we have two disjoint subchains. As is seen in figure~\ref{fig5} (a) and (b), the OTOC curves for various values of $\Delta$ for fixed $(i,j)$ pair with time, has a slope which increases with decreasing $\Delta$, whereas the information onset time is proportional to $\Delta$. The tunneling rate here basically refers to the junction flip rate which is extremely slow for higher values of $\Delta$. For every finite \(\Delta\) considered, the OTOC eventually reaches a late-time saturation regime. The higher the value of $\Delta$, larger is the time required for thermalization. This picture is akin to an intuitive classical picture of two containers, one filled with liquid and the other empty, connected by a pipe of variable thickness. The thickness of the pipe is analogous to $1/\Delta$, which controls the rate of flow of liquid from the filled container to the other. Irrespective of the pipe thickness, liquid in both the containers reach equilibrium heights akin to thermalization here. The next question is what is the timescale over which two parts of the chain can avoid bulk thermalization? Also, what is the timescale over which information insulation between junctions is relevant? There are in fact two characteristic timescales in the soft-junction case. The information onset time $t^{*}_{o} \sim \Delta$ as seen from figure~\ref{fig5}(b) signifies the time up to which the source operator information has not yet propagated past the junction and is below the numerical nonzero threshold. Therefore, the dynamics remains effectively indistinguishable from that of the hard-junction limit up to \(t_o^*\). From Appendix A, we see that the perturbative coupling between the $01$ and $10$ is $|J_{\mathrm{eff}}|=2\epsilon^{2}/\Delta$ for $\Delta \gg \epsilon$ which controls the rate of the virtual processes. This produces an associated coherent timescale \(t_{\rm eff}\sim1/|J_{\rm eff}|\sim\Delta/\epsilon^2\), consistent with the numerically observed scaling \(t_o^*\propto\Delta\) at fixed \(\epsilon\). The other timescale is  $t^{*}_{\Delta}$ which signifies the time range in which the subchains get increasingly entangled before reaching the equilibrium saturation value. Larger the value of $\Delta$, larger is this prethermal window.  As is seen in Fig~\ref{fig5}(c), the junction entanglement entropy is indistinguishable within resolution from the hard-junction case, across the cut made at the junction bond for $t\sim t^{*}_{o}$ . It eventually rises and reaches its saturation value near $t^{*}_{\Delta}$ which is an increasing function of $\Delta$ and the functional dependence is sensitive to the initial state that we start from. Here, we start from the vacuum state. In figure~\ref{fig5}(d), we see the generic increasing trend of the timescale $t^{*}_{\Delta}$ with $\Delta$ from the plot of entanglement entropy $S_{J}$, when the cut is made at the junction bond, as a function of time $t$ for the vacuum state. While the behaviour of the subchains and hence information insulation is practically indistinguishable within resolution from the hard-junction case for $t < t^{*}_{o}$, we see that the subchains do not maximally entangle when $t^{*}_{o}< t < t^{*}_{\Delta}$.

 \begin{figure}[ht!]
    \centering
    \includegraphics[width=\linewidth,height=5.5cm]{ 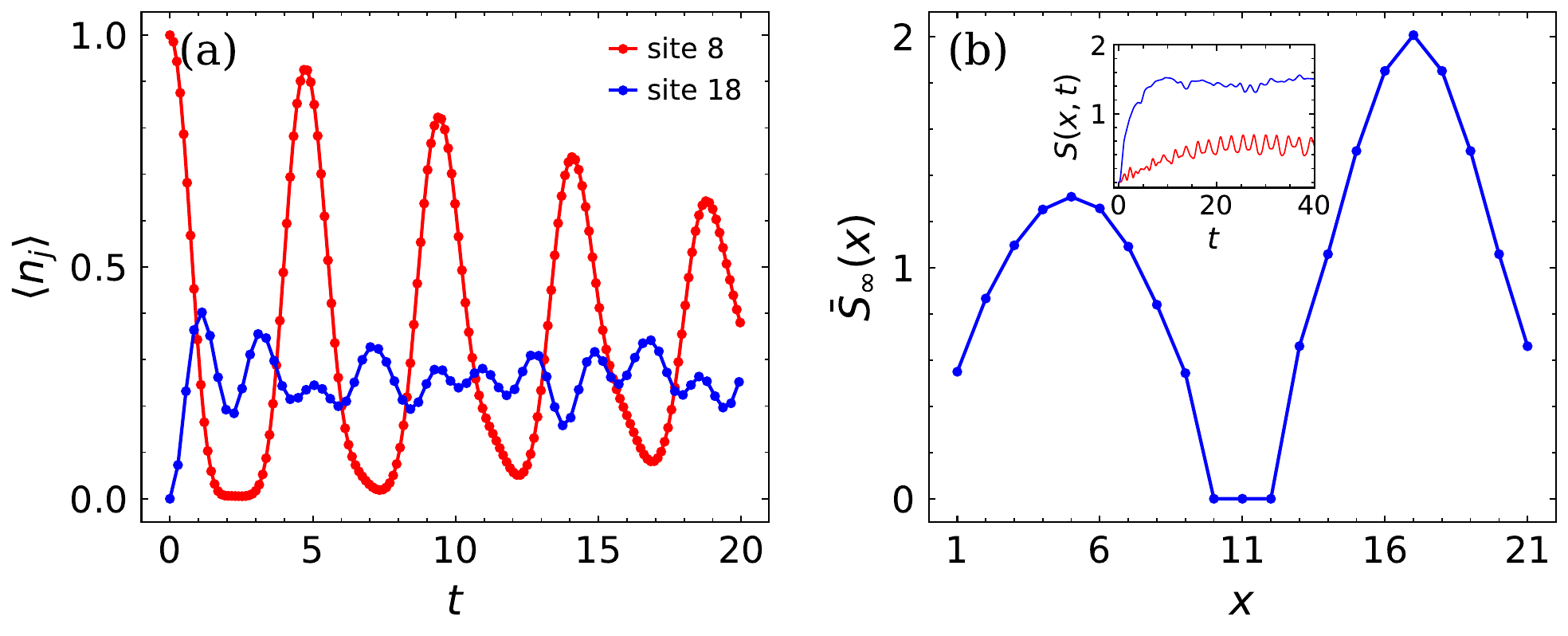}
    \includegraphics[width=\linewidth,height=5.7cm]{ 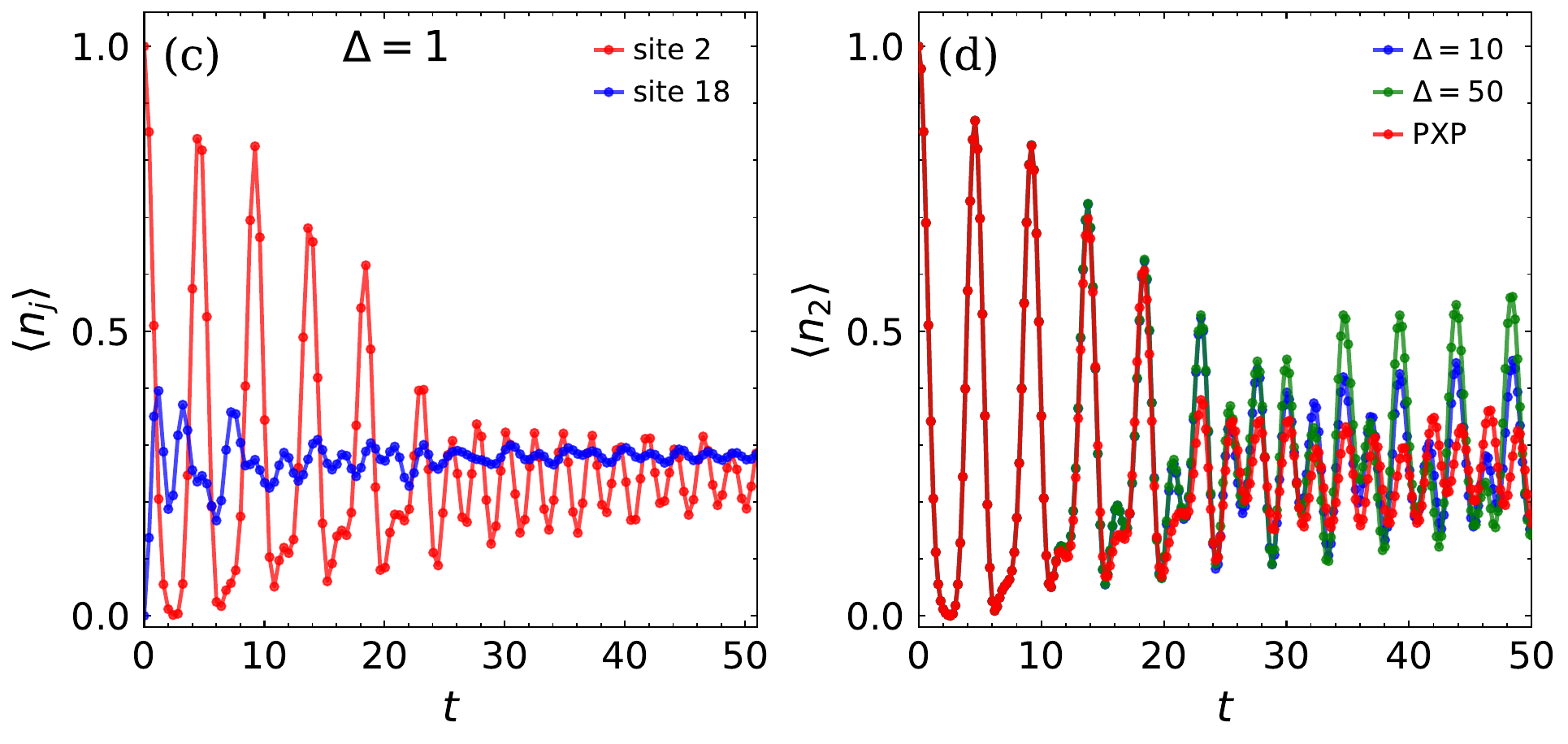}
  \caption{Panel $(a)$ shows the dynamics of the onsite density $\langle n_{r,l}^{j}(t)\rangle$ for two representative points, $j=8$ and $j=18$ in the two regions separated by the junction for a $PXP-d$-$PXP$ chain in the initial state $|01..01|01|11..11\rangle (L=22)$. Here, no $00$ or $11$ are allowed at the junction. Panel $(b)$ shows  the late-time average of the entanglement entropy, $\bar{S}_{\infty}(x)$ as a function of cuts along the chain where the cuts, $x$ are made on different bonds. Time variation of $S(x,t)$ is shown in the inset for $x=8$ and $x=18$. Panel $(c)$ shows the same quantities as in panel $(a)$ for sites $j=2$ and $j=18$, only now it is for the soft-junction with $\Delta=1$. The oscillations on the $Z_{2}$ half decay gradually over the prethermal window and ultimately thermalize to an $ETH$ value different from that of the vacuum half of the chain. In panel $(d)$, we plot the onsite density $\langle n_{l}^{j}(t)\rangle$ for the homogeneous $PXP$ chain with the blockade constraint and for the soft-junction $PXP-d$-$PXP$  chain with $\Delta=10, 50$ for the site $j=2$. In both cases, the initial states are $Z_{2}-$ vacuum for the respective constrained architectures for $L=22$. Increasing the value of $\Delta$ sustains a greater amplitude of oscillations for a longer interval as opposed to a much quicker decay in oscillations in case of the homogeneous $PXP$ chain. }
    {\label{fig6}}
\end{figure}   

 Another interesting feature of this model is that, in the hard-junction limit, we can create a state which partially thermalizes and partially does not thermalize as we go along the chain length. Suppose, we have a state with a $01$ junction $|01..01|01|11..11\rangle $ which has a $Z_{2}$ ordering in the left and a non-scarred Fock configuration on the right. We take two representative sites from  these two regions and plot the dynamics of their density $\langle n_{r,l}^{j}(t)\rangle$ as the Fock state evolves unitarily, in figure~\ref{fig6}(a). While the site in the $Z_{2}$ region undergoes long-lived revival oscillation, the dynamics of the other site rapidly thermalizes apart from some faint oscillations originating from finite-size effects of reflection from the edges. The frozen junctions keep the thermal and athermal regions disjoint and this is a novel example of what we would call a spatially heterogeneous dynamical state with scar-supported revivals in one half. In figure~\ref{fig6}(b), we plot the late-time average of the entanglement entropy $\bar{S}_{\infty}$ as a function of the cut on different bonds of this system which shows that the entanglement entropy in the left is much lower than that on the right. It is to be noted that the entanglement entropy in the $Z_{2}$ region keeps showing small oscillations, so we have plotted the average value at late-times. In the soft-junction scenario, for $t < t^{*}_{o}$, the gradient in the thermal properties across the junction is preserved like in the hard-junction case. Between $t^{*}_{o}< t < t^{*}_{\Delta}$, the subchains get gradually entangled and approach bulk thermalization. Thermalization is not a discontinuous step event but occurs gradually within the prethermal window. Therefore, disjoint thermal behaviour is far from ideality and the subchains can look appreciably dissimilar only at early and intermediate times  for appreciably high values of $\Delta$, in this time window. However, perfect insulation within numerical precision is still relevant for $t\sim \Delta$ where the junction behaves like a hard-junction. In figure~\ref{fig6}(c), we plot the excitation density $\langle n_{j}^{l,r}(t)\rangle$ for two-sites on the blockade and anti-blockade regions for the $Z_{2}-$vacuum initial state for $\Delta=1$. We see that although initially the oscillations persist on the $Z_{2}$ half, they decay away with time due to the soft-junction separating it from the vacuum and finally reach its thermal value. This is in contrast with figure~\ref{fig6}(a), where the $Z_{2}$ oscillations are completely insulated from the thermalizing half by the hard-junction. But since standard $PXP$ scar dynamics is associated with an approximate spectrum-generating \(SU(2)\)-like structure rather than an exact closed algebra~\cite{Choi2019}, there is a very slow decay in the oscillations even in case of the hard-junction, not owing to the properties of the junction. The oscillations are much more pronounced in amplitude in figure~\ref{fig6}(a) as compared to figure~\ref{fig6}(c) and this is because as we progressively increase \(\Delta\), the scarred segment approaches a shorter effective length; the shortest being realized in the hard-junction limit, and owing to these finite-size effects, the amplitude of the oscillations is reinforced. We also highlight here an observation that the junction and the open boundaries result in a system which is not translationally invariant and therefore, the $ETH$ values of $\langle n_{j}^{l,r}\rangle$ are in general site-dependent. This can be explained analytically by giving the simple example of an open boundary $PXP$ chain. The Hamiltonian in this case has $E\leftrightarrow-E$ symmetry which implies that a non-scarred initial state with zero average energy, should thermalize to an infinite temperature ensemble and therefore the boundary conditions in the Hamiltonian do not enter through the nontrivial Boltzmann weights in the thermal average of local observables. In fact, they enter through the constrained Hilbert space structure as well as due to the presence of the open boundaries. For example, here $\langle n_j\rangle_{\beta=0}=\frac{\operatorname{Tr}_{\mathcal H_{\rm con}}n_j}
{\dim\mathcal H_{\rm con}}
$ which implies $\langle n_{j}\rangle_{OBC}=\langle n_{L-j+1}\rangle_{OBC}=F_{j}F_{L-j+1}/F_{L+2}$. This  clearly is a site-dependent quantity. Here, $F_{j}$ represents the numbers from the Fibonacci sequence with $F_{1}=F_{2}=1$. For an unconstrained spin-\(\frac12\) Hilbert space at \(\beta=0\), each site has \(\langle n_j\rangle=1/2\), independent of position; the spatial variation in the $OBC$ $PXP$ chain is therefore purely a consequence of constrained-state counting and the presence of boundaries. It should be noted that the $PXP$ model with periodic boundary conditions has site independent densities even in the presence of the constrained architecture. Therefore, boundaries play a nontrivial role. In our model the \(E\leftrightarrow-E\) symmetry is absent, so a zero energy initial state is no longer generically tied to \(\beta_{\rm eff}=0\). Consequently, site dependence can arise both from constrained-state counting, open boundaries and from the nontrivial Boltzmann weighting of the spatially inhomogeneous Hamiltonian. In figure~\ref{fig6}(c), the densities on the two-sites reach visibly different thermal values at late times. Further, in~\ref{fig6}(d), we plot the excitation densities for a homogeneous $PXP$ chain and that for blockade-antiblockade structure with $\Delta=10,50$, on the $Z_{2}$ half of the $Z_{2}-$ vacuum initial state. Here, we clearly see that the $PXP$ oscillations die off faster than the soft-junction case; $\Delta=50$ being better at insulation than $\Delta=10$, as is evident from the figure.

\section{Experimental Realization}

This model can be realized in Rydberg atom arrays with the help of optical tweezers, addressing beams and facilitation techniques~\cite{Expt1,Expt2,Expt3,Expt4,Expt5}. The Rydberg Hamiltonian for the left and right chains containing $L/2$ sites each can be written as,

\begin{equation}
\mathcal{H}_{Ryd}=\Omega\sum_{i=1}^{L}\sigma_{x}^{i}-\sum_{i=1}^{L}\delta_{i}n_{i}+V_{NN}\sum_{i=1}^{L-1}n_{i}n_{i+1}
\end{equation}

where, $\Omega$ is the Rabi frequency of flipping between the Rydberg ground and excited states, i.e., $|g\rangle \leftrightarrow |e\rangle$, $\delta_{i}$ is the local detuning parameter and $V_{NN}$ is the nearest-neighbor van der Waals interaction between the Rydberg excited states. We create a spatially inhomogeneous detuning profile on the left and right as compared to the global flipping frequency $\Omega$. The left half of the chain should have a detuning, $\delta_{L}=0$ as compared to the global flipping laser frequency while the right side, through facilitation techniques, should be detuned to twice the nearest-neighbor interaction of the Rydberg states i.e., $\delta_{R}=2V_{NN}$ as in facilitated \(QXQ\)-type Rydberg dynamics~\cite{Zhao2025}. This spatial profile creates a frozen $01$ junction if we evolve a valid initial state with such a junction configuration while $10$ junction initial states are not frozen under this spatial detuning profile. In the following points, we explain how this protocol works.

\begin{itemize}
    \item For the left chain, since $\delta_{L}=0$, the cost for having $|g\rangle \equiv |0\rangle$ at the $i$-th site with $(0_{i-1},0_{i+1})$,$(0_{i-1},1_{i+1})$, $(1_{i-1},0_{i+1})$ or $(1_{i-1},1_{i+1})$ as nearest neighbors are all zero. If $|0\rangle \rightarrow |1\rangle$, then the energy cost for the previous set of neighbors become $0,V_{NN}, V_{NN}, 2V_{NN}$ respectively. Therefore, for $V_{NN}\gg \Omega$, apart from $0_{i-1}0_{i}0_{i+1}$, all the other flipping processes are off-resonant. Similarly, if we have $|r\rangle \equiv |1\rangle$ at the i-th site, $0_{i-1}1_{i}0_{i+1}$ initial configuration will only be resonant with the flipping $|1\rangle \rightarrow |0\rangle$ . So, starting from some {\it valid} initial state, the left side cannot have $11$ on neighboring sites which is the standard blockade regime.

    \item For the right chain, since $\delta_{R}=2V_{NN}$, the cost for having $|g\rangle \equiv |0\rangle$ at the $i$-th site with $(0_{i-1},0_{i+1})$,$(0_{i-1},1_{i+1})$, $(1_{i-1},0_{i+1})$ or $(1_{i-1},1_{i+1})$ as nearest neighbors are all zero. If $|0\rangle \rightarrow |1\rangle$, then the energy cost for the previous set of neighbors become $-2V_{NN},-V_{NN}, -V_{NN}, 0$ respectively. Therefore, for $V_{NN}\gg \Omega$, apart from $1_{i-1}0_{i}1_{i+1}$, all the other flipping processes are off-resonant. Similarly, if we have $|r\rangle \equiv |1\rangle$ at the i-th site, $1_{i-1}1_{i}1_{i+1}$ initial configuration will only be resonant with the flipping $|1\rangle \rightarrow |0\rangle$ . So, starting from some {\it valid} initial state, the right side cannot have $00$ on neighboring sites which is the standard anti-blockade regime.

    \item Now if we focus on the $0_{L}1_{R}$ junction configuration initial states, then $0_{L}$ has already a right neighbor which is $1_{R}$, therefore since it belongs to the left side which is the usual blockade regime, it will be frozen irrespective of what the other neighbor is. Similarly, if you look at $1_{R}$, then it already has a $0_{L}$ neighbor, so irrespective of what the other neighbor is, it will not be able to flip to $0$, since it is on the anti-blockade side. So, all Fock states intervened by $0_{L}1_{R}$ junction mimic the hard-junction limit.

    \item However, if we focus on the $1_{L}0_{R}$ junction configuration initial states, then $1_{L}$ can flip if its left side neighbor is also $0$, like its right side junction neighbor. Similarly, $0_{R}$ can flip if its right neighbor is also $1$, like its left junction neighbor. So, the junction is not frozen under this protocol and fails to mimic the hard-junction limit.

\end{itemize}

Hence, if we prepare an initial state like $|10..00|01|11..11\rangle$ where we create an excitation on the left half (with the help of addressing beams), then we will be able to observe the caging effect of information on the left side because as a consequence of the spatially different detunings, the junction remains automatically frozen while the left and right chains perform disjoint unitary evolution. It is also possible to engineer a $Z_{2}-$ vacuum state like $|01..01|01|11..11\rangle$ and observe how they evolve unitarily because the junction in this specific case is also frozen under the protocol defined earlier. This protocol mimics a hard-junction wall for a class of Fock states.

It is to be noted that a $PXP-PXP$ chain instead of a $PXP-d$-$PXP$ chain with the constraint that $00$ and $11$ are not allowed on the junction sites or the complementary constraint that $01$ and $10$ are not allowed on the junction sites, would show similar phenomena of information insulation and would also host partially athermal states. The internal symmetry in the latter case is $R$ for equal subchain $PXP-PXP$ junction, instead of $S$ as in the $PXP-d$-$PXP$ chain. It is particularly viable to engineer the soft-junction between two blockade regions, the reason behind which will be explained a little later.  If we write the diagonal two-site junction Hamiltonian in the Rydberg language then it has a direct correspondence with $\Delta_{00,11}$. The diagonal terms are $-\delta_{J}(n_{L/2}+n_{L/2+1})$ and $V_{J}(r)n_{L/2}n_{L/2+1}$. The cost of $00,11,01,10$ junction configurations are therefore $0,-2\delta_{J}+V_{J},-\delta_{J},-\delta_{J}$ respectively. We can shift all the energies by $\delta_{J}$ such that $E_{10,01}=0$ and $E_{00}=\delta_{J}$, $E_{11}=V_{J}-\delta_{J}$. Therefore, $\delta_{J}=\Delta_{00}$ and $V_{J}=\Delta_{11}+\Delta_{00}$. For the case, $\Delta_{00}=\Delta_{11}=\Delta$, $\delta_{J}=\Delta$ and $V_{J}=2\Delta$. The junction detuning can be  tuned by tuning the frequency of the local addressing beams whereas the interaction between the Rydberg atoms $V_{J}(r)$ can be tuned by tuning the separation $r$ between the atoms. Also since the Hamiltonian on the junction is an Ising-like term, it can be implemented by engineered control of the two-site terms in the Hamiltonian~\cite{Expt6,Expt7}. In the limit $\Delta\rightarrow \infty$, the soft-junction goes to the hard-constraint limit, but this limit is experimentally inaccessible since both the achievable local detuning and the Rydberg interaction strength are finite. Therefore, we can probe information insulation and disjoint thermal behaviour in the soft paradigm with appreciably large values of these parameters, which would give us access to a long prethermal window. The reason why we emphasized the $PXP-PXP$ chain here is because symmetric detunings at the junction sites are not compatible with anti-blockade constraint on the boundary bond neighboring to the right junction site, while there is no such experimental hindrance for the junction bridging two blockade regions. If we try to make the detunings on two junction sites different to be consistent with the anti-blockade constraint on the right boundary bond then, $E_{01}$ and  $E_{10}$ become nondegenerate, which is inconsistent with our model where $\{01,10\}$ form the degenerate low-energy manifold. This problem does not arise in the blockade regime thereby the choice of $PXP-PXP$ chain is being made for observing the exotic physics that soft-junctions render. Experimentally here facilitation becomes redundant but it is to be noted that our earlier model of $PXP-d$-$PXP$ chain is more natural when we think of dual constraints at the junction being born out of overlapping blockade and anti-blockade regions, but experimentally it is a little challenging to enforce the spatially varying constraint architecture.


\section{Conclusion}

We have, at first, proposed a novel model which can show Hilbert space fragmentation with the number of disconnected sectors growing exponentially with the number of domain wall/defect junctions in the chain. This fragmentation leads to a breakdown of global ETH and hence to ergodicity breaking. Consequently, this Hilbert space structure results in dynamical shielding of information by freezing the junction sites. We also show that it is possible to engineer partially athermal states where the junctions keep the thermal and athermal regions disjoint. The phenomenon of Hilbert space fragmentation is robust even when we have unequal-length $PXP$ and $d-PXP$ subchains separated by hard junctions. The number of Krylov fragments still scales as $2^n$, where $n$ is the number of hard junctions. Although here some global symmetries like $S=XR$ are generally lost. Secondly, we explore a more general regime where the hard constraints on the junction are softened by tuning a diagonal Hamiltonian parameter which decides the energy penalty for the previously forbidden states at the junction. We have shown here that information can be insulated and gradients in thermal behaviour can be preserved within numerical precision till a large onset time and at least partially over a long prethermal timescale when the previosuly projected junction states are penalised appreciably. This enables us to observe the rich physics of information insulation for an extended timescale without having to reach the strict hard-constraint limit. However, the Hilbert space fragmentation that we observe in the ideal infinite barrier limit is lost if we soften the junction. The level statistics evolves from predominantly Wigner-Dyson behavior at finite  \(\Delta\) toward Poisson-like statistics as the hard-junction limit is approached, with finite-size non-monotonicity; the exactly factorized $\Delta=0$ point being Poissonian. Moreover, there can be chirally protected zero-modes, pinned at zero energy by an exact symmetry \(\{\Gamma,H\}=0\), whose density profiles exhibit peaks near the physical edges and junctions. Also, as we have shown in the hard-constraint limit, there exist linear combinations in the zero-mode subspace that have anomalously low entanglement entropy and behave like zero-mode scars. These features make Fock states with large overlap with the protected zero-mode subspace promising for long-lived information retention. Finally, we have elaborately demonstrated how programmable Rydberg atom arrays present an exciting platform for realizing these novel phenomena in laboratories. It is also worthwhile to mention here that junctions can be used as barriers to isolate a subsystem from remote boundary environments. An environment can be modeled as reservoirs connected to the ends of the chain with different temperatures. This temperature gradient can induce thermal currents along the chain, to which the hard junction can act as an ideal barrier, while the soft junction can approximately insulate the region, with some leakage, for a timescale controlled by \(\Delta\). Therefore, junctions may be used to dynamically decouple a part of the system from environmental influence transmitted through the chain. A detailed study of this open-system setting is left for future work.

\section*{Acknowledgements}
 $AC$ would like to thank Arnab Sen for carefully reading the first version of the manuscript. $AC$ also thanks Krishnendu Sengupta for useful discussions. $AP$ would like to thank Abhijit Bandyopadhyay and Bobby Ezhuthachan for valuable discussions.

 \section*{Appendix A}
\label{app:SW}

In this Appendix, we derive the low-energy effective Hamiltonian in the regime $
\Delta_{00},\Delta_{11}\gg \epsilon_{00},\epsilon_{11}.
$ The low-energy junction manifold is
$
\mathcal{P}
=
\mathrm{span}\{|01\rangle,|10\rangle\},
$
whereas the high-energy manifold is
$
\mathcal{Q}
=
\mathrm{span}\{|00\rangle,|11\rangle\}.
$

We write the junction Hamiltonian in the form
\begin{equation}
H=H_0+V,
\end{equation}
where
\begin{equation*}
H_0
=
\Delta_{00}|00\rangle\langle00|
+
\Delta_{11}|11\rangle\langle11|,
\end{equation*}
and
\begin{align*}
V={}&
\epsilon_{00}
\Big[
P_l^{L/2-1}|10\rangle\langle00|
+
P_r^{L/2+2}|00\rangle\langle01|
+\mathrm{H.c.}
\Big]
\nonumber\\
&+
\epsilon_{11}
\Big[
P_r^{L/2+2}|10\rangle\langle11|
+
P_l^{L/2-1}|11\rangle\langle01|
+\mathrm{H.c.}
\Big].
\end{align*}

It is convenient to define
\begin{align*}
T_{00}
&=
\epsilon_{00}|00\rangle
\left(
\langle10|P_l^{L/2-1}
+
\langle01|P_r^{L/2+2}
\right),
\\
T_{11}
&=
\epsilon_{11}|11\rangle
\left(
\langle10|P_r^{L/2+2}
+
\langle01|P_l^{L/2-1}
\right),
\end{align*}
such that
\begin{equation*}
V
=
T_{00}+T_{00}^{\dagger}
+
T_{11}+T_{11}^{\dagger}.
\end{equation*}

The Schrieffer-Wolff transformation is generated by an anti-Hermitian operator $S$,
\begin{equation}
\widetilde H
=
e^SHe^{-S},
\qquad
S^\dagger=-S.
\end{equation}
Expanding the transformed Hamiltonian using the Baker-Campbell-Hausdorff series, we get
\begin{equation}
\widetilde H
=
H+[S,H]
+
\frac12[S,[S,H]]
+\cdots.
\end{equation}
The operator $S$ is chosen so that the coupling between the low- and high-energy manifolds vanishes to first order:
\begin{equation*}
V+[S,H_0]=0.
\end{equation*}
Since
\begin{equation*}
[T_{\alpha},H_0]
=
-\Delta_{\alpha}T_{\alpha},
\qquad
[T_{\alpha}^{\dagger},H_0]
=
\Delta_{\alpha}T_{\alpha}^{\dagger},
\qquad
\alpha=00,11,
\end{equation*}
the required generator is
\begin{equation}
S
=
\frac{T_{00}-T_{00}^{\dagger}}{\Delta_{00}}
+
\frac{T_{11}-T_{11}^{\dagger}}{\Delta_{11}}.
\end{equation}

Retaining terms up to order $\epsilon^2/\Delta$, one obtains
\begin{align*}
\widetilde H
={}&
H_0+V
+[S,H_0]
+[S,V]
+
\frac12[S,[S,H_0]]
+
\mathcal{O}
\left(
\frac{\epsilon^3}{\Delta^2}
\right).
\end{align*}
Using
\begin{equation*}
[S,H_0]=-V,
\end{equation*}
we have
\begin{equation*}
\frac12[S,[S,H_0]]
=
-\frac12[S,V],
\end{equation*}
and therefore
\begin{equation}
\widetilde H
=
H_0
+
\frac12[S,V]
+
\mathcal{O}
\left(
\frac{\epsilon^3}{\Delta^2}
\right).
\end{equation}

In order to get the low energy effective Hamiltonian, we now project onto the low-energy manifold using
\begin{equation*}
P
=
P_{01}+P_{10}
=
|01\rangle\langle01|
+
|10\rangle\langle10|.
\end{equation*}
The corresponding low-energy effective Hamiltonian is
\begin{equation}
H_{\mathrm{eff}}
=
P\widetilde H P.
\end{equation}
Since
\begin{equation*}
PH_0P=0,
\end{equation*}
the leading correction is
\begin{equation*}
H_{\mathrm{eff}}^{}
=
\frac12
P[S,V]P.
\end{equation*}
Using the explicit form of $S$ and $V$, this reduces to
\begin{equation}
H_{\mathrm{eff}}^{}
=
-
\frac{T_{00}^{\dagger}T_{00}}{\Delta_{00}}
-
\frac{T_{11}^{\dagger}T_{11}}{\Delta_{11}}.
\end{equation}

The products appearing above are
\begin{align*}
T_{00}^{\dagger}T_{00}
=
\epsilon_{00}^{2}
\Big[
&
P_l^{L/2-1}|10\rangle\langle10|
+
P_r^{L/2+2}|01\rangle\langle01|
+
P_l^{L/2-1}P_r^{L/2+2}
\left(
|10\rangle\langle01|
+
|01\rangle\langle10|
\right)
\Big],
\end{align*}
and
\begin{align*}
T_{11}^{\dagger}T_{11}
=
\epsilon_{11}^{2}
\Big[
&
P_r^{L/2+2}|10\rangle\langle10|
+
P_l^{L/2-1}|01\rangle\langle01|
+
P_l^{L/2-1}P_r^{L/2+2}
\left(
|10\rangle\langle01|
+
|01\rangle\langle10|
\right)
\Big].
\end{align*}

Hence the second-order junction contribution is
\begin{align}
H_{\mathrm{eff}}^{}
={}&
-
\left[
\frac{\epsilon_{00}^{2}}{\Delta_{00}}
P_l^{L/2-1}
+
\frac{\epsilon_{11}^{2}}{\Delta_{11}}
P_r^{L/2+2}
\right]
|10\rangle\langle10|
-
\left[
\frac{\epsilon_{00}^{2}}{\Delta_{00}}
P_r^{L/2+2}
+
\frac{\epsilon_{11}^{2}}{\Delta_{11}}
P_l^{L/2-1}
\right]
|01\rangle\langle01|
\nonumber\\
&-
P_l^{L/2-1}P_r^{L/2+2}
\left(
\frac{\epsilon_{00}^{2}}{\Delta_{00}}
+
\frac{\epsilon_{11}^{2}}{\Delta_{11}}
\right)
\times
\left(
|10\rangle\langle01|
+
|01\rangle\langle10|
\right).
\end{align}

The full low-energy many-body Hamiltonian is therefore
\begin{equation}
H_{\mathrm{low}}
=
H_{01}\oplus H_{10}
+
H_{\mathrm{eff}}.
\end{equation}

For the symmetric case, $\Delta_{00}
=
\Delta_{11}
=
\Delta$,
$
\epsilon_{00}
=
\epsilon_{11}
=
\epsilon,$ the perturbatively generated coupling between the $01$ and $10$ sectors becomes
\begin{equation}
J_{\mathrm{eff}}
=
-
\frac{
2\epsilon^{2}
}{\Delta}
P_l^{L/2-1}P_r^{L/2+2}.
\end{equation}

Thus, for any finite but large $\Delta$, the two low-energy sectors are weakly coupled through virtual excursions into the $00$ and $11$ manifolds. In the strict limit
$\Delta\rightarrow\infty,$
all second-order corrections vanish as $1/\Delta$, and therefore
\begin{equation}
H_{\mathrm{low}}
\longrightarrow
H_{01}\oplus H_{10}.
\end{equation}
Hence the direct-sum structure of the hard-junction is recovered in the infinite-detuning limit.

\bibliography{SciPost_Example_BiBTeX_File.bib}

\end{document}